\newcommand{\bra}[1]{\ensuremath{\left\langle {#1} \right|}}
\newcommand{\ket}[1]{\ensuremath{\left|  #1 \right\rangle}}
\newcommand{\expec}[1]{\ensuremath{\left\langle {#1} \right\rangle}}

\newcommand{\Goz}{\ensuremath{W\,}}

\newcommand{\Jperp}{\ensuremath{J^2_{\perp}}}

\newcommand{\W}{\ensuremath{W}}
\newcommand{\Wp}{\ensuremath{2\gamma_\perp}}
\newcommand{\Wpk}{\ensuremath{W_{\mathrm{opt}}}}
\newcommand{\Gt}{\ensuremath{\Gamma_{3e}}}

\newcommand{\ddt}{\ensuremath{\frac{\mathrm{d}}{\mathrm{d}t}}}

\newcommand*\xbar[1]{%
  \hbox{%
    \vbox{%
      \hrule height 0.5pt 
      \kern0.4ex
      \hbox{%
        \kern-0.1em
        \ensuremath{#1}%
        \kern-0.1em
      }%
    }%
  }%
} 

\documentclass[aps,prl,amsmath,amssymb,reprint,notitlepage,showpacs]{revtex4-1}
\usepackage{ulem}
\usepackage{amsmath}
\usepackage{color}
\usepackage{graphicx}
\usepackage{hyperref}
\usepackage{calligra}
\DeclareMathOperator{\Tr}{Tr}
\DeclareMathAlphabet{\pazocal}{OT1}{pzc}{m}{it}
\DeclareMathAlphabet{\mathcalli}{T1}{calligra}{m}{n}
\DeclareFontShape{T1}{calligra}{m}{n}{<->s*[2]callig15}{}

\begin{document}
\title{Linear response theory for superradiant lasers}
\author{Justin G. Bohnet}
\author{Zilong Chen}
\author{Joshua M. Weiner}
\author{Kevin C. Cox}
\author{James K. Thompson}
\affiliation{JILA, NIST and Department of Physics, University of Colorado, Boulder, Colorado 80309-0440, USA }
\date{\today}


\begin{abstract}

We theoretically study a superradiant laser, deriving both the steady-state behaviors and small-amplitude responses of the laser's atomic inversion, atomic polarization, and light field amplitude. Our minimum model for a three-level laser includes atomic population accumulating outside of the lasing transition and dynamics of the atomic population distribution causing cavity frequency tuning, as can occur in realistic experimental systems.  We show that the population dynamics can act as real-time feedback to stabilize or de-stabilize the laser's output power, and we derive the cavity frequency tuning for a Raman laser. We extend the minimal model to describe a cold-atom Raman laser using $^{87}$Rb, showing that the minimal model qualitatively captures the essential features of the more complex system \cite{BCWDyn}. This work informs our understanding of the stability of proposed millihertz linewidth lasers based on ultranarrow optical atomic transitions and will guide the design and development of these next-generation optical frequency references.

\end{abstract}

\maketitle

\section{I. Introduction}
\label{sec:intro}

Steady-state, superradiant lasers based on narrow optical atomic transitions have the potential to be highly stable optical frequency references, with unprecedentedly narrow quantum-limited linewidths below 1~millihertz \cite{MYC09,CHE09,BCW12}.  These lasers may achieve such high frequency stability because the laser linewidth and frequency are determined primarily by the atomic transition rather than the cavity properties.  As a result, the lasing frequency is predicted to be many orders of magnitude less sensitive to both the thermal and technical mirror motion that currently limits the frequency stability of passive optical reference cavities \cite{kessler2012sub, Hinkley2013}. The insensitivity to vibration means that superradiant lasers may be able to stably operate outside carefully engineered, low-vibration laboratory environments for both practical and fundamental applications \cite{LTB11, APL12}. 

To minimize inhomogeneous broadening of the atomic transition, proposed narrow-linewidth superradiant lasers would use trapped, laser-cooled atoms as the gain medium \cite{MYC09,CHE09}. The first use of cold atoms as a laser gain medium was reported in Ref. \cite{HFG92}. Recently, the spectral properties of a cold-atom Raman laser were studied in a high finesse cavity, deep into the so-called good-cavity regime \cite{PhysRevLett.107.063904}.  Clouds of cold atoms can also simultaneously provide gain and feedback for distributed feedback lasing \cite{SZC12} and random lasing \cite{Baudouin2013}. Cold atoms have also been used as the gain medium in four-wave mixing experiments \cite{Greenberg2012, BGB10, PhysRevLett.91.203001} and in collective atomic recoil lasing \cite{Kruse2003}. 

Beyond the technical applications, superradiant lasers are of fundamental interest. The narrow natural and inhomogenous linewidths provided by laser trapped and cooled atoms means that proposed superradiant lasers are bad-cavity lasers. This unusual regime of laser physics is accessed when the cavity linewidth is much larger than the linewidth of the gain medium. 
The quantum-limited linewidth of a bad-cavity laser follows the Schawlow-Townes linewidth \cite{SCT58} usually applied to microwave masers \cite{Haken84, PhysRevLett.72.3815, MYC09}, instead of the linewidth applied to optical lasers that typically operate in the opposite good-cavity limit.  Bad-cavity lasers near the cross-over regime (i.e. where the cavity linewidth is approximately equal to the linewidth of the gain medium) have yielded signatures of chaos, demonstrating the predicted equivalence to the Lorenz model \cite{Haken197577,WB86}. 
Operation of a laser deep into the bad-cavity regime has only recently begun to be studied in detail using laser cooled atoms as the gain medium \cite{BCW12,BCWDyn,WCB12,BCWHybrid}. The first analysis of a cold-atom, superradiant Raman laser focused on the laser phase noise \cite{BCW12}. 

This paper presents theoretical studies of both the steady-state and amplitude stability properties of a superradiant laser. Our work both guides the future implementation of proposed superradiant optical lasers, and explains already experimentally-realized superradiant Raman lasers.  This work directly supports the experimental efforts using Raman transitions in $^{87}$Rb of Refs. \cite{BCW12, BCWDyn, BCWHybrid, WCB12}. 

The key new results presented here are a simple minimum model that nonetheless captures the qualitative features in recent experimental demonstrations, a derivation of crucial laser emission frequency tuning effects in a good or bad cavity cold-atom Raman laser, and an investigation of laser amplitude instabilities caused by frequency tuning effects in the case of a bad-cavity laser. 

We begin in Sec. II by constructing a model of a steady-state Raman laser that makes three extensions to the two-level superradiant laser model presented in Ref. \cite{MYC09}. These extensions are motivated by pumping and cavity tuning effects present in the experimental work of Refs. \cite{BCW12, BCWDyn, BCWHybrid, WCB12}.  The extensions include: (1) an imperfect atomic repumping scheme in which some population remains in an intermediate third level, (2) additional decoherence caused by Rayleigh scattering during pumping, and (3) a tuning of the cavity mode frequency in response to the distribution of atomic populations among the available atomic states. The model makes no assumptions about operation in the good or bad cavity regime. 

\begin{figure}[t]
\includegraphics[width=3.3in]{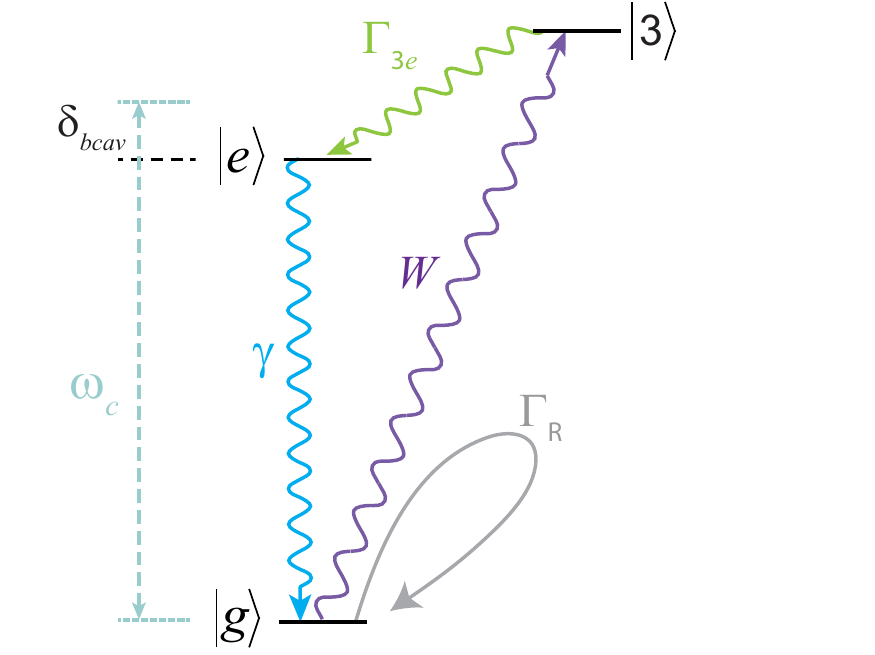}
\caption{(color online) Energy level diagram of a three-level superradiant laser using the optical transition from \ket{e} to \ket{g}. The emitted optical laser light (blue) is nearly resonant with the cavity mode (dashed lines), detuned from $\omega_{\mathrm{c}}$ by $\delta_{bcav}$. The atoms are incoherently pumped to a third state at a rate $W$.  Atoms in \ket{g} also Rayleigh scatter at a rate $\Gamma_R$, but leaves them in state \ket{g}.  The incoherent pumping from \ket{3} to \ket{e} at rate $\Gamma_{3e}$ completes the cycle.
}
\label{3lvl}
\end{figure}

We then restrict ourselves to considering only bad-cavity regime and linearize the coupled atom-field equations about steady-state to study the small signal response of the laser to external perturbations. 
We identify relaxation oscillations and dynamic cavity feedback that can serve to damp or enhance oscillatory behavior of the laser amplitude.

In Sec. III, we explicitly show that a Raman lasing transition involving three levels can be reduced to the previous section's two level lasing transition. 
The formalism directly produces the cavity tuning in response to atomic populations that was introduced by hand in Section II.  

Finally, in Sec. IV, we model the experimental $^{87}$Rb Raman system of Refs. [3,21], incorporating all eight atomic ground hyperfine states. 
We derive the steady-state behavior and linear response to small perturbations of this more realistic system and compare the qualitative features to the results of the more simple model introduced in Sec. II and III.

\section{II. Three-Level Model}

\subsection{A. Deriving the laser equations}

We begin by equations for a general three-level laser, making no assumptions about a good cavity or bad cavity regime. The three-level model for the laser presented here is pictured in Fig. \ref{3lvl}.  It consists of two lasing levels denoted by excited state $\ket{e}$ and ground state \ket{g} separated by optical frequency $\omega_{eg}$, a third state \ket{3} which the atoms must be optically pumped to before they can be optically pumped back to \ket{e}, and a single optical cavity mode with resonance frequency $\omega_c$.  The cavity resonance is near the $\ket{e}\rightarrow\ket{g}$ transition frequency, with $\delta_{bcav} = \omega_c - \omega_{eg}$.  We describe the atoms-cavity system using the Jaynes-Cummings  Hamiltonian \cite{jaynes1963comparison}

\begin{equation}
\hat{H} = \hbar \omega_c \hat{c}^\dag \hat{c} + \hbar g (\hat{c}^\dag \hat{J}_- + \hat{c}\hat{J}_+).
\label{eqn:H}
\end{equation}

\noindent Here $2g$ is the single atom vacuum Rabi frequency that describes the strength of the coupling of the atoms to the cavity mode, set by the atomic dipole matrix element.  The operators $\hat{c}$ and $\hat{c}^\dag$ are the bosonic annihilation and creation operators for photons in the cavity mode.   We have introduced the collective spin operators $\hat{J}_- = \sum^N_{q=1} \ket{g^{(q)}}\bra{e^{(q)}}$, and $\hat{J}_+ = \sum^N_{q=1} \ket{e^{(q)}}\bra{g^{(q)}}$ for the \ket{e} to \ket{g} transition, assuming uniform coupling to the cavity for each atom. The index $q$ labels the sum over $N$ individual atoms. We also define the number operator for atoms in the state \ket{k}, $k\in\{e, g, 3 \}$, as $\hat{N}_k = \sum^N_{q=1}\ket{k^{(q)}}\bra{k^{(q)}}$ and the collective spin projection operator $\hat{J}_z = \frac{1}{2} \sum^N_{q=1} ( \hat{N}^{(q)}_e - \hat{N}^{(q)}_g)$. 

The density matrix for the atom cavity system is $\hat{\rho} = \sum_{kl}\sum^\infty_{m,n=0}\ket{k,n}\bra{l,m}$ where the second sum is over the atomic basis states $k,l \in \{g,e,3\}$, and the third sum is over the cavity field basis of Fock states. The time evolution of $\hat{\rho}$ is determined by a master equation for the atom cavity system 

\begin{equation}
\dot{\hat{\rho}} = \frac{1}{i\hbar} [\hat{H},\hat{\rho}] + \mathcal{L}[\hat{\rho}]. 
\label{eqn:masterEquation}
\end{equation}

Dissipation is introduced through the Liouvillian $\mathcal{L}[\hat{\rho}]$ \cite{MYC09}.   Sources of dissipation and associated characteristic rates include the power decay rate of the cavity mode at rate $\kappa$,  the spontaneous decay from \ket{e} to \ket{g} at rate $\gamma$, the spontaneous decay from \ket{3} to \ket{e} at rate $\Gt$, and Rayleigh scattering from state $\ket{g}$ at rate $\Gamma_R$. The repumping, usually just called pumping in other laser literature, is treated as ``spontaneous absorption" at rate $W$, analogous to spontaneous decay, but from a lower to higher energy level. Physically, this is achieved by coupling \ket{g} to a very short lived excited state that decays to \ket{3}.  The Liouvillian is written as a sum of contributions from the processes above respectively as  $\mathcal{L}[\hat{\rho}]=\mathcal{L}_c[\hat{\rho}] + \mathcal{L}_{eg}[\hat{\rho}] + \mathcal{L}_{3e}[\hat{\rho}] + \mathcal{L}_{R}[\hat{\rho}] + \mathcal{L}_{g3}[\hat{\rho}]$. The individual Liouvillians are given in Appendix A.

%
%
%
%

\begin{figure}
\includegraphics[width=2.5in]{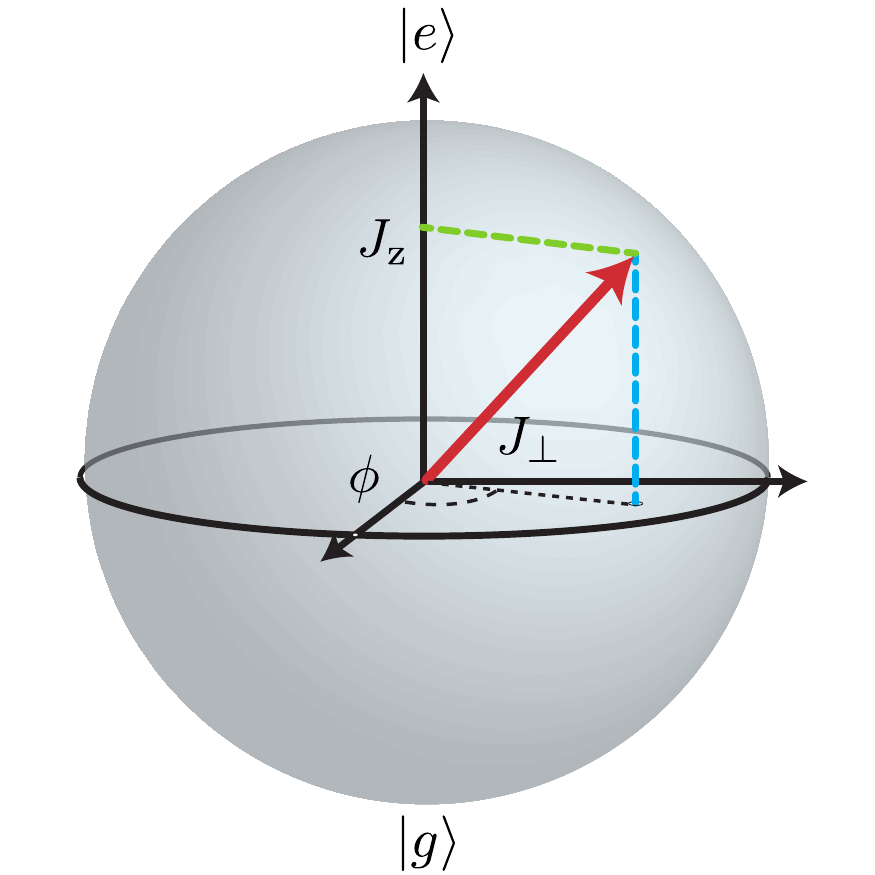}
\caption{Representing the state of the \ket{e}, \ket{g} two level system using a collective Bloch vector.}
\label{fig:blochvector}
\end{figure}

We obtain equations of motion for the relevant expectation values of the atomic and field operators using $\dot{O} = \Tr[\hat{O}\dot{\hat{\rho}}]$.  Complex expectation values are indicated with script notation, while real definite expectation values are standard font, so $\dot{\mathcal{C}} = \Tr[\hat{c}\dot{\hat{\rho}}]$ and $\dot{\mathcal{J}}_- = \Tr[\hat{J_-}\dot{\hat{\rho}}]$. We assume the unknown emitted light frequency is $\omega_\gamma$ and factor this frequency from the expectation values for the cavity field and the atomic polarization, $\mathcal{C} = \breve{\mathcal{C}}e^{-i\omega_\gamma t}$ and $\mathcal{J}_- = \breve{\mathcal{J}}_-e^{-i\omega_\gamma t}$.  The symbol $\breve{\, \,}$ indicates a quantity in a frame rotating at the laser frequency.  The set of coupled atom-field equations is then

\begin{gather}
\dot{\breve{\mathcal{C}}} = -(\kappa/2 + i (\omega_c - \omega_\gamma))\breve{\mathcal{C}} - i g \breve{\mathcal{J}}_- \label{eqn:2lvlcavity}\\
\dot{\breve{\mathcal{J}}}_- = -(\gamma_\perp + i(\omega_{eg} - \omega_\gamma)) \breve{\mathcal{J}}_- + i 2g \breve{\mathcal{C}} J_z \label{eqn:2lvlJm}\\
\begin{split}
\dot{J}_z =-\left(W+\gamma  \right)\frac{J_z}{2} + \left(2\Gt-\W+\gamma\right)\frac{N_3}{4} \\
  +  \frac{N}{4}\left(W-\gamma \right)  + i g (\breve{\mathcal{J}}_- \breve{\mathcal{C}}^* + \breve{\mathcal{J}}_+ \breve{\mathcal{C}})
\label{eqn:2lvldotJz}
\end{split}\\
\dot{N}_3 = - \left(\Gt + \W/2 \right) N_3 + \W \left(N/2 - J_z\right)
\label{eqn:N3TwoLvlDiffEq}
\end{gather}

In the above equations, we have combined the broadening of the atomic transition into a single transverse decay $\gamma_\perp =\gamma/2 + W/2 + \Gamma_R/2$. We have assumed no entanglement between the atomic degrees of freedom and the cavity field in order to factorize expectation values of the form $\expec{\hat{\sigma}_{kl} \hat{c}}=\expec{\hat{\sigma}_{kl}} \expec{\hat{c}}$. The equations make no assumptions about the relative sizes of the various rates, making them general equations for a three level laser, but one of the distinct differences in cold-atom lasers versus typical lasers is that the transverse decay rate is often dominated by the repumping rate $\gamma \sim W/2$.

It is useful to represent the two-level system formed by \ket{e} and \ket{g} as a collective Bloch vector (Fig. \ref{fig:blochvector}).  The vertical projection of the Bloch vector is given by the value of $J_z$, and is proportional to the laser inversion. The projection of the Bloch vector onto the equatorial plane $J_\perp$ is given by the magnitude of atomic polarization $|\breve{\mathcal{J}}_-|$, with $J^2_\perp = |\breve{\mathcal{J}}_-|^2$.  We refer to $J_\perp$ as the collective transverse coherence of the atomic ensemble.


\subsection{B. Steady-state solutions}
To understand how extending to this three-level model affects the fundamental operation of the laser, we now study the steady-state solutions with respect to repumping rates, cavity detuning, and Rayleigh scattering rates. The steady-state solutions assume $\gamma \ll W$, the regime of operation for proposed superradiant light sources \cite{MYC09,CHE09} and the experiments of Refs. \cite{BCW12, BCWDyn}, but make no approximations based on the relative magnitudes of $\gamma_\perp$ and $\kappa$. Thus $\gamma_\perp \approx W/2 + \Gamma_R/2$, but otherwise the results in the section hold for both good-cavity ($\kappa \ll \gamma_\perp$)  and bad-cavity ($\kappa \gg \gamma_\perp$)  lasers.  We first determine the steady-state oscillation frequency, starting by setting the time derivatives in Eqn. \ref{eqn:2lvlcavity} and Eqn. \ref{eqn:2lvlJm} to zero.  After solving for~ $\breve{\mathcal{C}},$ 

\begin{equation}
\breve{\mathcal{C}} = - i \frac{g}{\kappa/2 + i \delta_0} \breve{\mathcal{J}}_-
\label{eqn:elimC}
\end{equation}

\noindent where $\delta_0$ denotes the cavity detuning from the laser emission frequency $\delta_0 = \omega_c - \omega_\gamma$. Substituting the result into Eqn. \ref{eqn:2lvlJm}, we have 

\begin{equation}
2g^2 J_z = (\gamma_\perp + i(\omega_{eg} -\omega_\gamma))(\kappa/2 + i \delta_0)\,\, .
\label{eqn:LaserFreqConstraint}
\end{equation}

\noindent Since $J_z$ is always real, the imaginary part of Eqn. \ref{eqn:LaserFreqConstraint} must be zero. This constrains the frequency of oscillation to

\begin{equation}
\omega_\gamma = \frac{\omega_c}{1+\frac{\kappa}{2\gamma_\perp}} + \frac{\omega_{eg}}{1+\frac{2\gamma_\perp}{\kappa}}\,\, ,
\end{equation}

\noindent a weighted average of the cavity frequency and the atomic transition frequency. 

We solve for the steady-state solutions of Eqns. \ref{eqn:2lvlcavity}-\ref{eqn:2lvldotJz} by setting the remaining time derivatives to zero and substituting Eqn. \ref{eqn:elimC} for $\mathcal{C}$ in all the equations. In this work, the amplitude properties are our primary interest (as compared to the phase properties studied in Refs. \cite{MYC09,BCW12}), so we further simplify the equations, at the expense of losing phase information, by considering the magnitude of the atomic polarization $|\breve{\mathcal{J}}_-|$.  The equation for the time derivative of $J_\perp^2$ is

\begin{gather}
 \frac{d}{dt}J^2_\perp = |\breve{\mathcal{J}}_+| |\frac{d}{dt} \breve{\mathcal{J}}_-| + |\breve{\mathcal{J}}_-| |\frac{d}{dt} \breve{\mathcal{J}}_+| \label{eqn:ddtJperpdef}
\end{gather}

The steady-state output photon flux is just proportional to the square of the equatorial projection

\begin{equation}
\dot{\bar{M}}_c = \kappa|\breve{\mathcal{C}}|^2 = J_\perp^2\frac{C\gamma}{1+\delta_0'^2} \,\, .
\end{equation} 

\noindent Here we have also defined a normalized detuning $\delta_0' =\delta_0/(\kappa/2)$ and a single particle cavity cooperativity parameter 

\begin{equation}
C = \frac{(2g)^2}{\kappa\gamma}
\label{eqn:singleparticleCparameter}
\end{equation}

\noindent that gives the ratio of single-particle decay rate from \ket{e} to \ket{g} for which the resulting photon is emitted into the cavity mode, making $C$ equivalent to the Purcell factor \cite{TanjiSuzuki2011201}.  

\begin{figure}
\includegraphics[width= 3in]{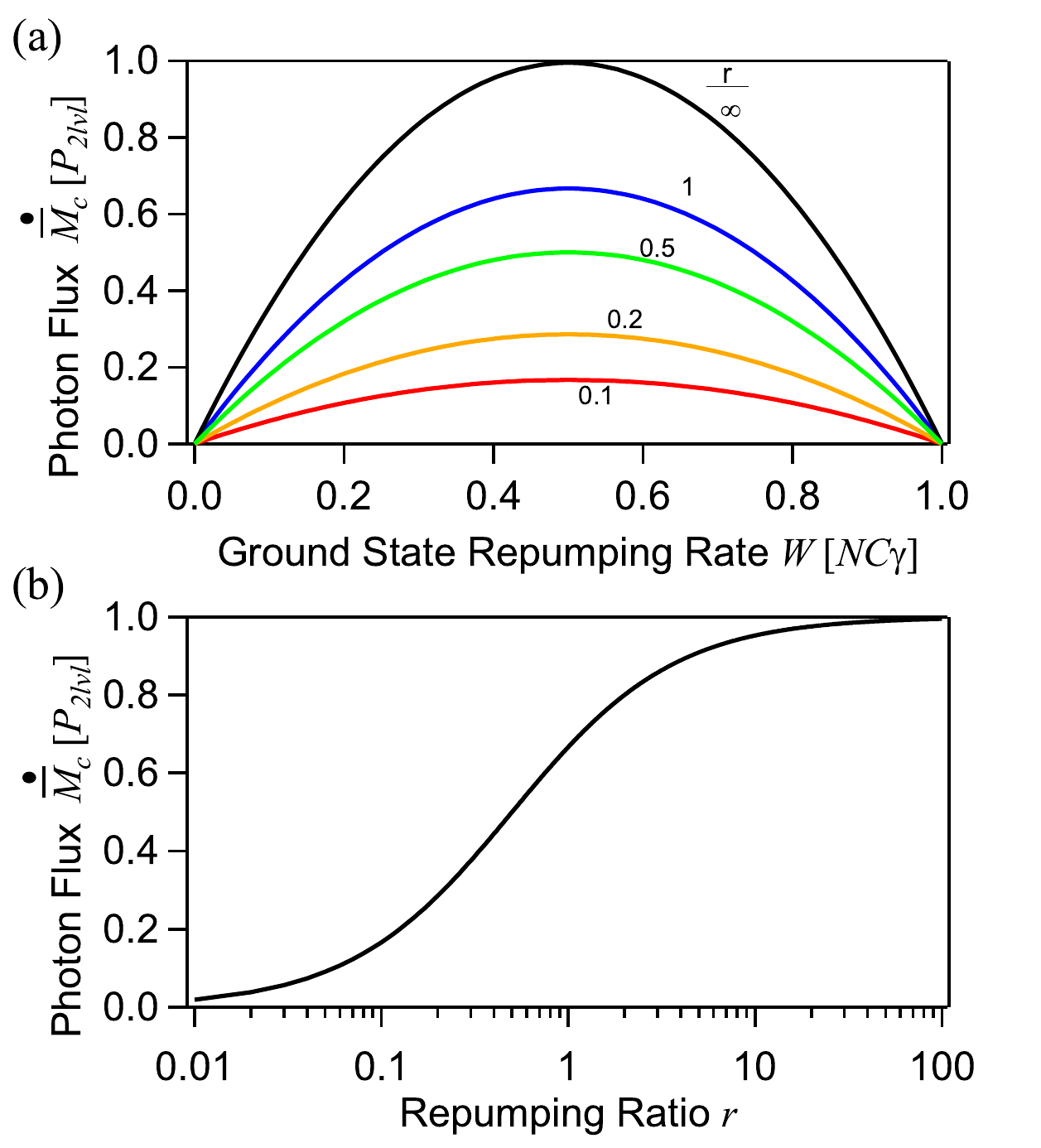}
\caption{(a) Steady-state photon flux $\dot{\bar{M}}_c$ versus ground state repumping rate $W$, with a series of curves showing the effects of repumping through the additional state \ket{3}.The case of $r = \infty$ is the two level model of Ref. \cite{MYC09} (black).  (b) Photon flux $\dot{\bar{M}}_c$ versus $r$ with $W = W_{opt}$. For all curves, $\delta_0' = 0$ and $\Gamma_R = 0$, and the photon flux is plotted in units of $P_{2lvl} = N^2C\gamma/8$.   }
\label{fig:fluxVsr}
\end{figure}

After these substitutions and simplifications, the steady-state solutions (denoted with a bar) are

\begin{equation}
\bar{J}_{z}= \frac{\Wp(1+\delta_0'^2)}{2 C\gamma}
\label{eqn:SSJz}
\end{equation}

\begin{align}
\begin{split}
\bar{J}^2_{\perp} = \left(\frac{N}{2}\right)^2 \left(\frac{2r}{1/2+r} \right) \left( \frac{ \W (1+\delta_0'^2)}{NC\gamma}\right) \left(1-\frac{\Wp(1+\delta_0'^2)}{NC\gamma}\right)
\label{eqn:SSJperp}
\end{split}
\end{align}

\begin{equation}
\bar{N}_{3}= \frac{\bar{J}^2_{\perp}} {r}\left(\frac{C\gamma}{ \W(1+\delta_0'^2)}\right)
\label{eqn:SSN3}
\end{equation}

\begin{equation}
\dot{\bar{M}}_c = \left(\frac{N}{2}\right)^2 \left(\frac{2r}{1/2+r} \right) \left(\frac{W}{N}\right) \left(1-\frac{\Wp(1+\delta_0'^2)}{NC\gamma}\right)
\label{eqn:McSS2lvl}
\end{equation} 

\noindent written in terms of the repumping ratio $r \equiv \Gamma_{3,e}/ W$. Note that $r$ also determines the steady-state build up of population in \ket{3} as $\bar{N}_3/\bar{N}_g = 1/r$.  To succinctly express the modification of $\bar{J}_\perp^2$ and $\dot{\bar{M}}_c$ due to inefficient repumping, we define the reduction factor

\begin{equation}
R(r) \equiv \frac{r}{1/2+r}\,\, .
\label{eqn:Rfactor3lvl}
\end{equation}

\noindent which appears in Eqns. \ref{eqn:SSJperp} and \ref{eqn:McSS2lvl} above.

\begin{figure}
\includegraphics[width= 3in]{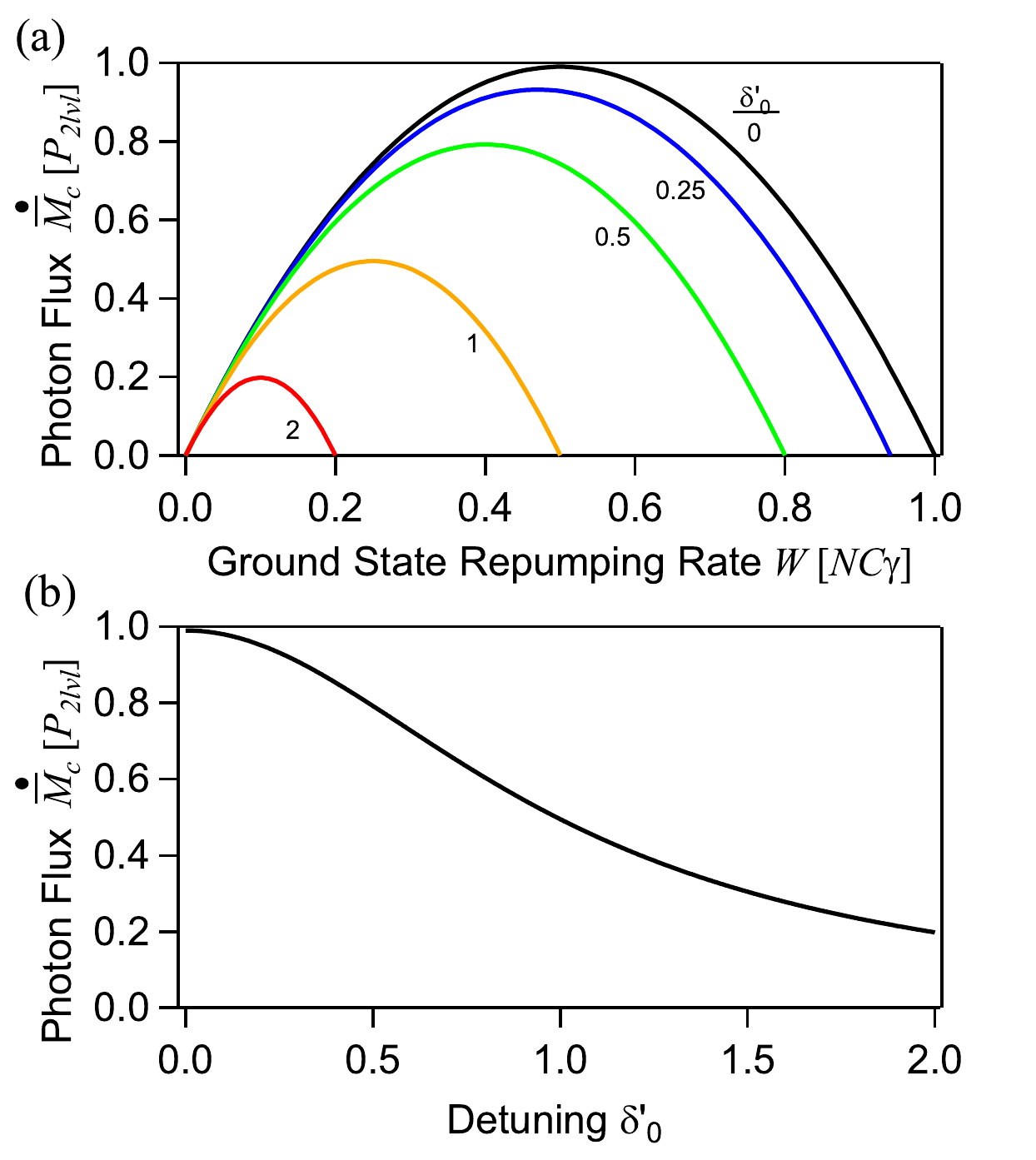}
\caption{(a) Steady-state photon flux $\dot{\bar{M}}_c$ versus ground state repumping rate $W$, with a series of curves showing the effects of detuning of the cavity resonance frequency from the emitted light frequency $\delta_0'$. (b) Photon flux $\dot{\bar{M}}_c$ versus $\delta_0'$ with $W = W_{opt}(\delta_0')$. The photon flux is plotted in units of $P_{2lvl} = N^2C\gamma/8$.  For all curves, $r = \infty$ and $\Gamma_R = 0$.}
\label{fig:fluxVsd}
\end{figure}
\begin{figure}
\includegraphics[width= 3in]{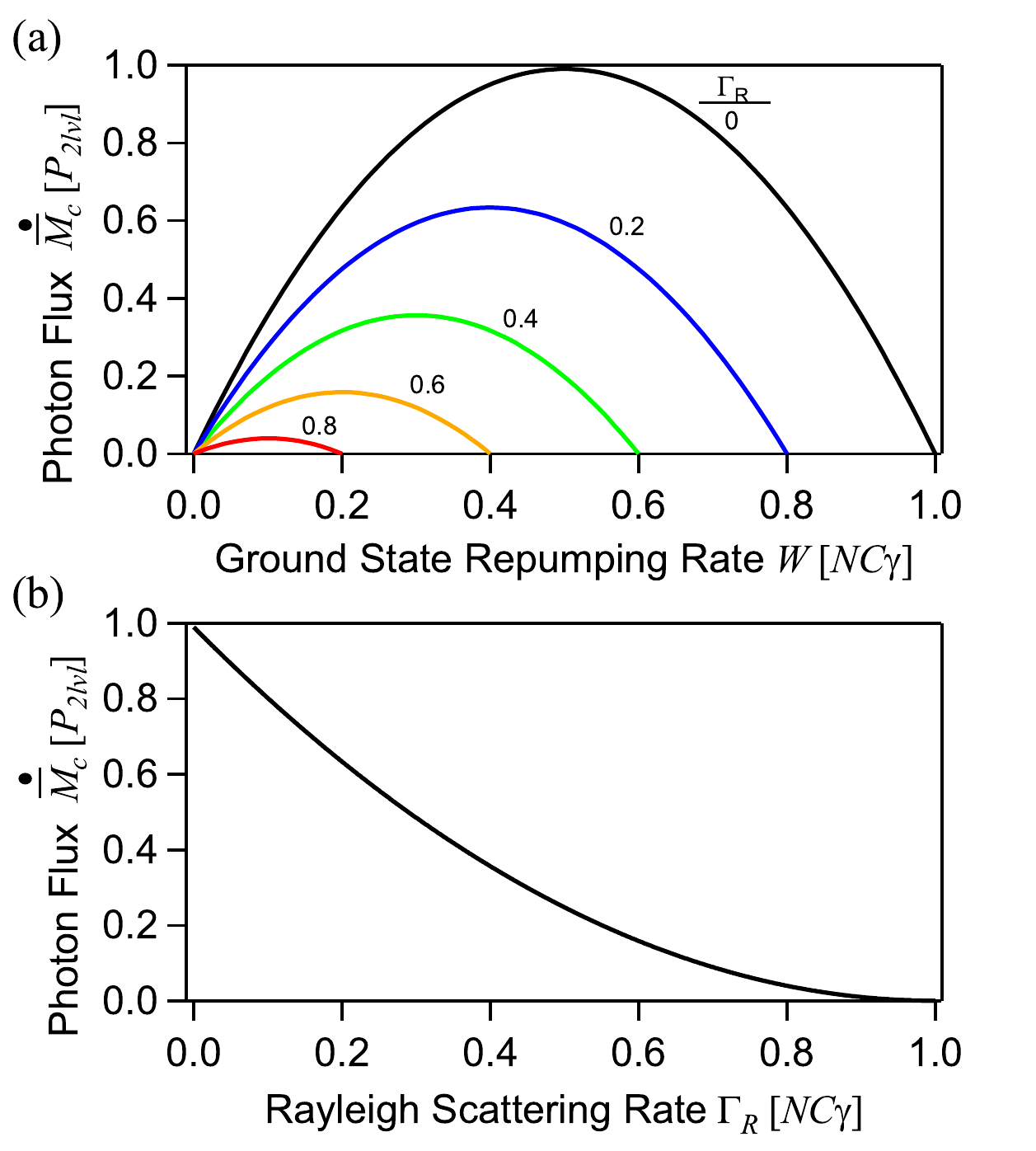}
\caption{Steady-state photon flux $\dot{\bar{M}}_c$ versus ground state repumping rate $W$, with a series of curves showing the effects of decoherence in the form of Rayleigh scattering from the ground state \ket{g}. (b) Photon flux $\dot{\bar{M}}_c$ versus $\Gamma_R$ with $W = W_{opt}(\Gamma_R)$. The photon flux is plotted in units of $P_{2lvl} = N^2C\gamma/8$. For all curves,  $r = \infty$ and $\delta_0' = 0$.}
\label{fig:fluxVsGR}
\end{figure}

Next we discuss the behavior of these solutions for the characteristic parameters of the three-level model: $W$, $r$, $\delta_0'$, and $\Gamma_R$. The results are illustrated in Figs. \ref{fig:fluxVsr}-\ref{fig:fluxVsGR}.  

First, we focus on the impact of repumping on the steady-state behavior.  The photon flux $\dot{\bar{M}}_c$ follows a parabolic curve versus the ground state repumping rate $W$ (Fig. \ref{fig:fluxVsr}).  In the limit $\delta_0' \rightarrow 0$, $r \rightarrow \infty$, and $\Gamma_R \rightarrow 0$, Eqn. \ref{eqn:McSS2lvl} reduces to the result for the simple two-level model of Ref. \cite{MYC09}.  This limit is shown as the black curve in part (a) of Figs. \ref{fig:fluxVsr}-\ref{fig:fluxVsGR}. At low $W$, the photon flux is limited by the rate at which the laser recycles atoms that have decayed to \ket{g} back to \ket{e}.  At high $W$, the photon flux becomes limited by the decoherence from the repumping, causing the output power to decrease with increasing $W$.  When the atomic coherence decays faster than the collective emission can re-establish it, the output power goes to zero. This decoherence limit is expressed in the condition for the maximum repumping threshold, above which lasing ceases:

\begin{equation}
W_{max} = \frac{NC\gamma}{1+\delta_0'^2} - \Gamma_R\,\, .
\end{equation}  

\noindent The output photon flux is optimized at $W_{opt} = W_{max}/2$.  Notice that the maximum repumping rate is not affected by $r$. However, the additional decoherence (here in the form of Rayleigh scattering) lowers the turn-off threshold. If $\Gamma_R > \frac{NC\gamma}{1+\delta_0'^2}$, the decoherence will prevent the laser from reaching superradiant threshold regardless of $W$. 

In Figs. \ref{fig:fluxVsr}-\ref{fig:fluxVsGR}, we plot Eqn. \ref{eqn:McSS2lvl} emphasizing (a) the modification to the photon flux parabola, and (b) the optimum photon flux as a function of the population in the third state (as parameterized by the repumping ratio $r$), detuning of the cavity resonance from the emission frequency $\delta$, and additional decoherence from Rayleigh scattering $\Gamma_R$. The photon flux is plotted in units of the optimum photon flux in the two-level model of Refs. \cite{MYC09, MEH10}, $P_{2lvl} = N^2C\gamma/8$.

As the repumping process becomes more inefficient and population builds up in \ket{3},  parameterized by $r$ as $\bar{N}_3/\bar{N}_g =1/r$, we see from Eqns. \ref{eqn:SSJperp} and \ref{eqn:McSS2lvl} that the photon flux $\dot{\bar{M}}_c$ decreases (Fig. \ref{fig:fluxVsr}). A repumping ratio $r=10$ ensures the laser operates within a few percent of its maximum output power. Notice that $\dot{\bar{M}}_c$ saturates after $r$ is greater than $\approx 2$. Although inefficient repumping suppresses $\dot{\bar{M}}_c$, the optimum and maximum repumping rates $W_{opt}$ and $W_{max}$ are not modified.

The preservation of the operating range can be important, as in practice large values of $r$ can lead to added decoherence (due to intense repumping lasers for example), which does reduce the operating range.  Lowering the value of $r$ allows some flexibility as some output power can be sacrificed to keep the laser operating over a wider range of $\xbar{W}$.

Cavity detuning modifies both the $\dot{\bar{M}}_c$ and $W_{\mathrm{opt}}$ (Fig. \ref{fig:fluxVsd}). The modification arises from the $\delta_0'$ dependent cavity cooperativity

\begin{equation}
C' = \frac{C}{1+\delta_0'^2}\,\, .
\end{equation}

\noindent The modified cooperativity $C'$ originates from the atomic polarization radiating light at $\omega_\gamma$, which non-resonantly drives the cavity mode with the usual Lorentzian-like frequency response. Thus, the output photon flux $\dot{\bar{M}}_c$, turn-off threshold $W_{max}$, and optimum repumping rate $W_{opt}$ all scale like $1/(1+\delta_0'^2)$. This effect is symmetric with respect to the sign of $\delta_0'$. Physically, the rate a single atom spontaneously decays from \ket{e} to \ket{g} by emitting a photon into the cavity mode is $\Gamma_c \equiv C'\gamma$, which we use to simplify some later expressions.

Finally, we examine the effect of additional atomic broadening through $\Gamma_R$ in Fig. \ref{fig:fluxVsGR}.  Additional broadening linearly reduces $W_{opt}$ and $W_{max}$, but because we require the repumping rate to remain at $W_{opt}$ in Fig \ref{fig:fluxVsGR}b, $\dot{\bar{M}}_c$ has a $(\frac{\Gamma_R}{N\Gamma_c}-1)^2$ dependence.

The key insight from the steady-state solutions for our three-level model is that imperfections in the lasing scheme can quickly add up, greatly reducing the expected output power of the laser. A repumping scheme should be chosen to minimize Rayleigh scattering $\Gamma_R$ and maximize the repumping ratio $r$.  Added decoherence, as well as the detuning $\delta_0'$ are especially problematic because they restrict the possible range of $W$ for continuous operation.   



\subsection{C. Linear expansion of uncoupled equations}

For future applications of steady-state superradiant light sources as precision measurement tools, we are interested in the system's robustness to external perturbations. As is common in laser theory\cite{M66, siegman86, PhysRevA.47.1431}, here we analyze the system's linear response to perturbations by considering small deviations from the steady-state solutions. While all previous expressions are valid for both the good-cavity and bad-cavity limit, as no assumptions were made about the relative magnitudes of $\kappa$ and $\gamma_\perp$, it is convenient now to simplify to two equations for the dynamics by assuming that the laser is operating deep in the bad-cavity regime, where $\kappa \gg \gamma_\perp$. In this regime, the cavity field adiabatically follows the atomic polarization, providing the physical motivation to eliminate the field from Eqns. \ref{eqn:2lvlcavity}-\ref{eqn:N3TwoLvlDiffEq} \cite{MYC09,PhysRevLett.72.3815}.



The cavity field is eliminated by assuming that the first time derivative of the complex field amplitude $\breve{\mathcal{C}}$ in Eqn. \ref{eqn:2lvlcavity} is negligible compared to $\frac{\kappa}{2}\mathcal{C}$. This effectively results in Eqn. \ref{eqn:elimC} being the equation for the cavity field. After substituting Eqn. \ref{eqn:elimC} into Eqns. \ref{eqn:2lvlJm}-\ref{eqn:N3TwoLvlDiffEq}, we only concern ourselves with the amplitude responses, simplifying the equations by using Eqn. \ref{eqn:ddtJperpdef} and substituting $|\mathcal{J}_-|^2$ with $J_\perp^2$. With these simplifications, the dynamical equations for $\dot{J}_z$, $\dot{J}^2_\perp$, and $\dot{N}_3$ are

\begin{equation}
\dot{J_z} =  ((\Gt-\W/2) \frac{N_3}{2} + \frac{W}{2}(N/2 - J_z)) - \frac{C \gamma}{1+\delta_0'^2} \Jperp
\label{eqn:Jz2lvl}
\end{equation}

\begin{equation}
\dot{\Jperp} = - \Wp \Jperp  + \frac{2 C \gamma}{1+\delta_0'^2} J_z \Jperp.
\label{eqn:Jperp2lvl}
\end{equation}

\begin{equation}
\dot{N}_3 = - \left(\Gt + \W/2 \right) N_3 + \W \left(N/2 - J_z\right)
\end{equation}


We perform the linear expansion by re-parameterizing the degrees of freedom in terms of fractionally small perturbations about steady-state: $J_z(t) = \bar{J}_{z}(1+{j}_z(t))$, $J^2_\perp(t) = \bar{J}^2_{\perp}(1+2{j}_\perp(t))$, and $N_3(t) = \bar{N}_{3}(1+{n}_3(t))$. We also define the response of cavity field through the relationship $A(t) \equiv \sqrt{|\mathcal{C}(t)|^2} = \bar{A}(1+a(t))$.  Since $|\mathcal{C}(t)|^2 = \frac{C\gamma}{1+\delta_0'(t)'^2} J^2_\perp(t)$ from Eqn. \ref{eqn:elimC}, $A(t)$ follows the atomic polarization, except for the modification from dynamic cavity detuning as will be discussed below. We analyze the response in the presence of a specific form of external perturbation -- the modulation of the repumping rate $W(t) = \xbar{W}(1+ w(t))$ with $w(t) = \epsilon\mathrm{Re}[e^{i\omega t}]$, where $\epsilon$ is a real number much less than 1. The quantities ${j}_z(t)$, ${j}_\perp(t)$, ${n}_3(t)$, $a(t)$, and $w(t)$ are unitless fractional perturbations around the steady-state values that we assume are much less than $1$. 

We also include, by hand, an inversion-dependent term in the detuning $\delta_0' = \delta' +\alpha \bar{J}_z {j}_z(t)$. The cavity mode's frequency is tuned by the presence of atoms coupled to the cavity mode.  The tuning is equal but opposite for atoms in the two different quantum states \ket{e} and \ket{g}.  The detuning $\delta'$ is the steady-state value of the detuning of the dressed cavity from the emitted light frequency. The variation about this steady-state detuning is governed by the second contribution $\alpha\bar{J}_z {j}_z(t)$.  Effects such as off-resonant dispersive shifts due to coupling to other states can lead to this $J_z$ dependent detuning in real experiments.  We derive this cavity tuning in Sec. III. 

To linearize the resulting equations, we substitute the expansions around steady-state into Eqns. \ref{eqn:N3TwoLvlDiffEq}, \ref{eqn:Jz2lvl}, and \ref{eqn:Jperp2lvl}. We neglect  terms beyond first order in the small quantities ${j}_z(t)$, ${j}_\perp(t)$, ${n}_3(t)$, $a(t)$, and $w(t)$. For ease of solving the equations, we treat ${j}_z(t)$, ${j}_\perp(t)$, ${n}_3(t)$, and $a(t)$ as complex numbers where the real part gives the physical value. After eliminating the steady-state part of the equations, the equations for small signal responses ${j}_\perp(t)$  and ${j}_z(t)$ can be reduced to two uncoupled, third order differential equations


\begin{gather}
\beta \dddot{\jmath}_{\perp}  + \ddot{\jmath}_{\perp} +  2 \gamma_0 \dot{\jmath}_{\perp} +  \omega_0^2  \jmath_{\perp} = D_{\perp}(\omega)\epsilon e^{i\omega t} \label{eqn:DiffEqResponse1}\\
\beta \dddot{\jmath}_{z}  + \ddot{\jmath}_{z} +  2 \gamma_0 \dot{\jmath}_{z} +  \omega_0^2  \jmath_{z} = D_{z}(\omega)\epsilon e^{i\omega t}.
\label{eqn:DiffEqResponse2}
\end{gather}

\noindent We have written the uncoupled differential equations in a form that suggests a driven harmonic oscillator, with damping rate $\gamma_0$, natural frequency $\omega_0$ and a drive unique to the $\jmath_\perp$ or $\jmath_z$ equation $D_{\perp}(\omega)$ or $D_{z}(\omega)$.  The drives contain derivatives of the repumping modulation $w(t)$, resulting in frequency dependence. The third derivative term is a modification to the harmonic oscillator response from the third level, characterized by the factor $\beta$ that goes to zero in the two-level limit ($r \rightarrow \infty$).  To preserve the readability of the text, we have included the full expressions for the coefficients as Appendix A.  Each of the terms will be discussed subsequently in physically illuminating limits.

The drive of this harmonic oscillator-like system varies with the modulation frequency and other system parameters.  In the case of the two-level model of Ref. \cite{MYC09}, with $r= \infty$, $\alpha = 0$, and $\Gamma_R = 0$, the drive terms are 

\begin{gather}
D_\perp(\omega) = \frac{\xbar{W}}{2}(N\Gamma_c-2\xbar{W}-i\omega) \label{eqn:D_perpSimple} \\
D_z(\omega) = (N\Gamma_c-\xbar{W})\xbar{W}+i\omega).
\end{gather}

The modulation-frequency-dependent terms add an extra 90$^\circ$ of phase shift at high modulation frequencies to the observed response.  Additionally, the cancellation in $D_\perp(\omega=0)$ results in an insensitivity of the output photon flux to the ground state repumping rate $W$ at $W_{opt}$. The cancellation agrees with the parabolic dependence of $\dot{\bar{M}}_c$ on $W$, as seen in the steady-state solutions.

The frequency dependent terms in $D_{\perp,z}$ also cause a growing drive magnitude versus $\omega$.  This is canceled out in the responses $j_{\perp}$ and $j_z$ by the roll-off from the oscillator, keeping the response finite versus modulation frequency. These characteristic features remain in the response, even as the complexity of the model increases as additional effects are included.  



To proceed, we solve the equations for the complex, steady-state response to a single modulation frequency $\omega$, (e.g. $\jmath_\perp(t) = \jmath_\perp e^{i\omega t}$). The complex response of the cavity field amplitude $a(t)$ results from these solutions,

\begin{equation}
a(t) = \jmath_\perp(t) - \frac{\delta \alpha \bar{J}_{z} \jmath_z(t)}{1+\delta'^2}\,\, .
\label{eqn:cavityfieldresponse}
\end{equation}

In contrast to Eqn. \ref{eqn:elimC}, where $|\mathcal{C}|$ depends only on $J_\perp$, including dispersive cavity tuning from the inversion couples the cavity output power to $J_z$ as well.

\subsection{D. Transfer function analysis} 

\begin{figure}
\includegraphics[width=3in]{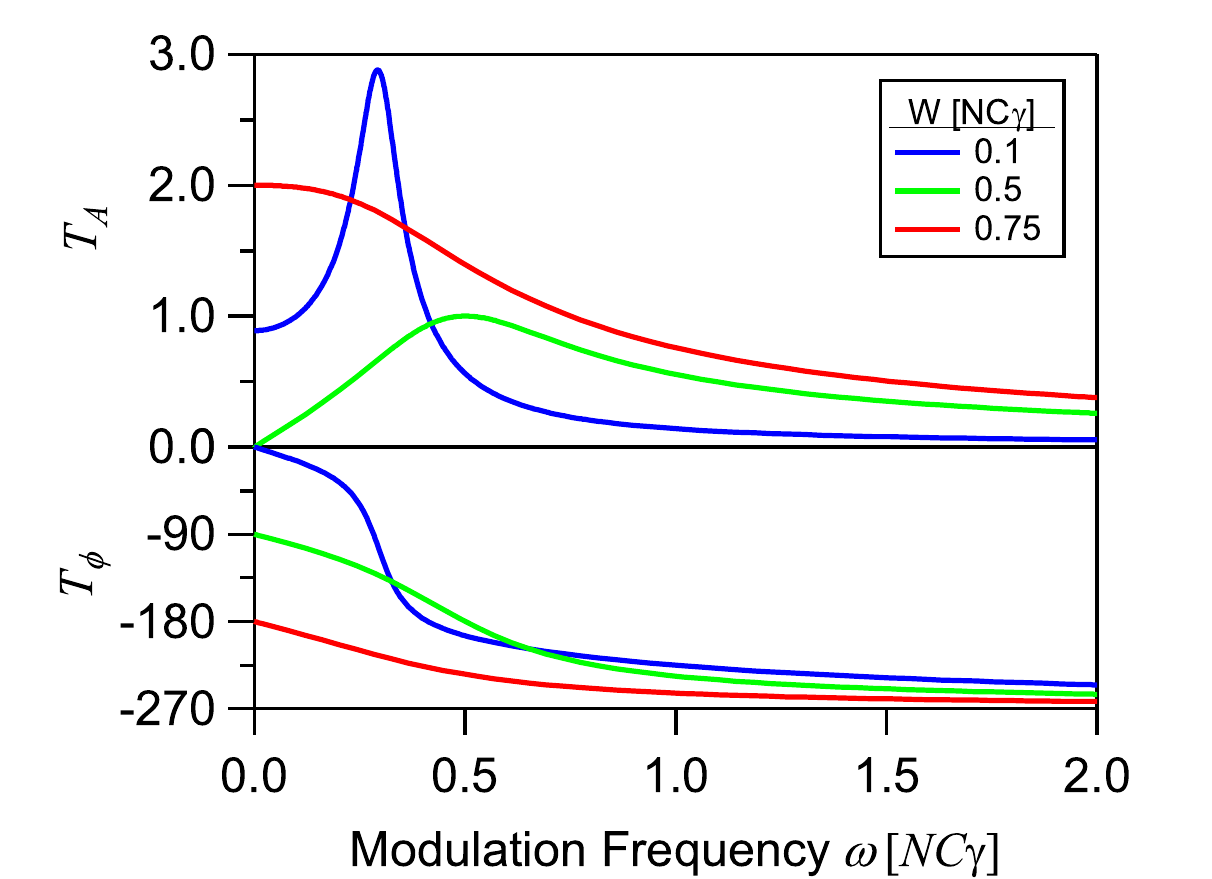}
\caption{Output photon flux transfer function for different ground state repumping rates, with $r = \infty$, $\delta' = 0$, $\alpha = 0$ and $\Gamma_R = 0$. }
\label{fig:TfuncVsW}
\end{figure}

\begin{figure}
\includegraphics[width=3in]{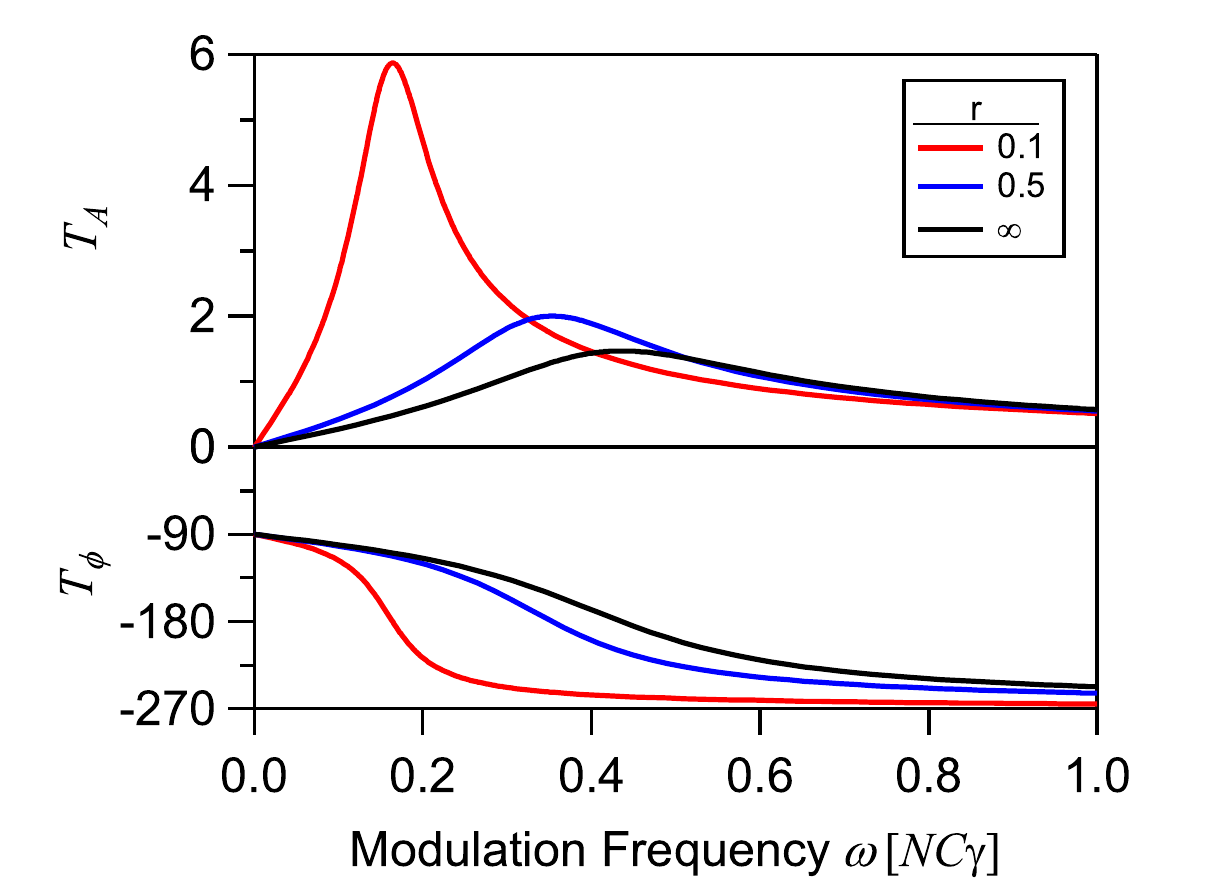}
\caption{Output photon flux transfer function for different repumping ratios $r$, with $W = W_{opt}$, $\delta' = 0$, $\alpha = 0$ and $\Gamma_R = 0$.}
\label{fig:TfuncVsr}
\end{figure}

\begin{figure}
\includegraphics[width=3in]{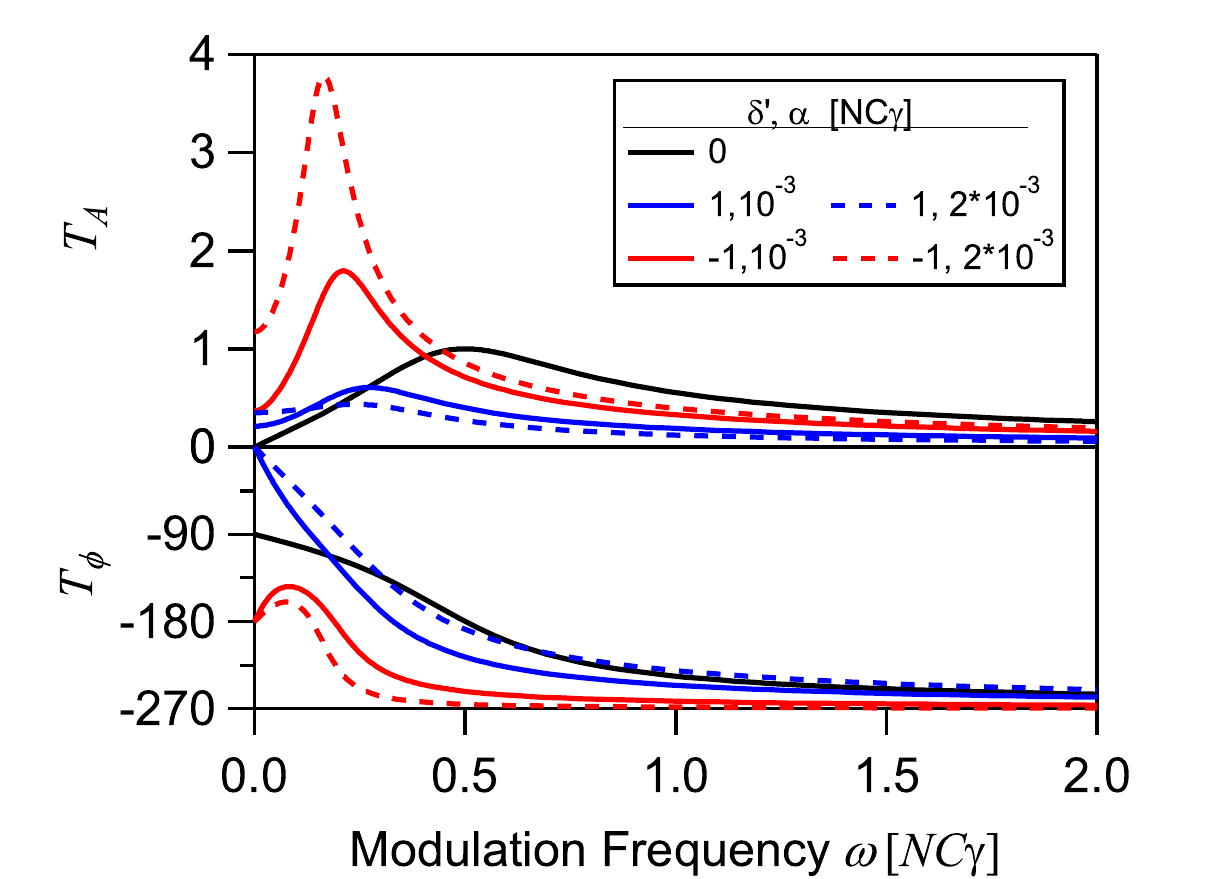}
\caption{Output photon flux transfer function for different dressed cavity detuning from emitted light frequency $\delta$, with $W = W_{opt}(\delta) = \frac{NC\gamma}{2(1+\delta'^2)}$, $r = \infty$, and $\Gamma_R = 0$. The solid (dashed) lines show $\alpha = NC\gamma \times 10^{-3}$ ($\alpha = NC\gamma \times 2 (10^{-3})$) to demonstrate the effects of increased cavity feedback.
}
\label{fig:TfuncVsd}
\end{figure}

\begin{figure}
\includegraphics[width=2.8in]{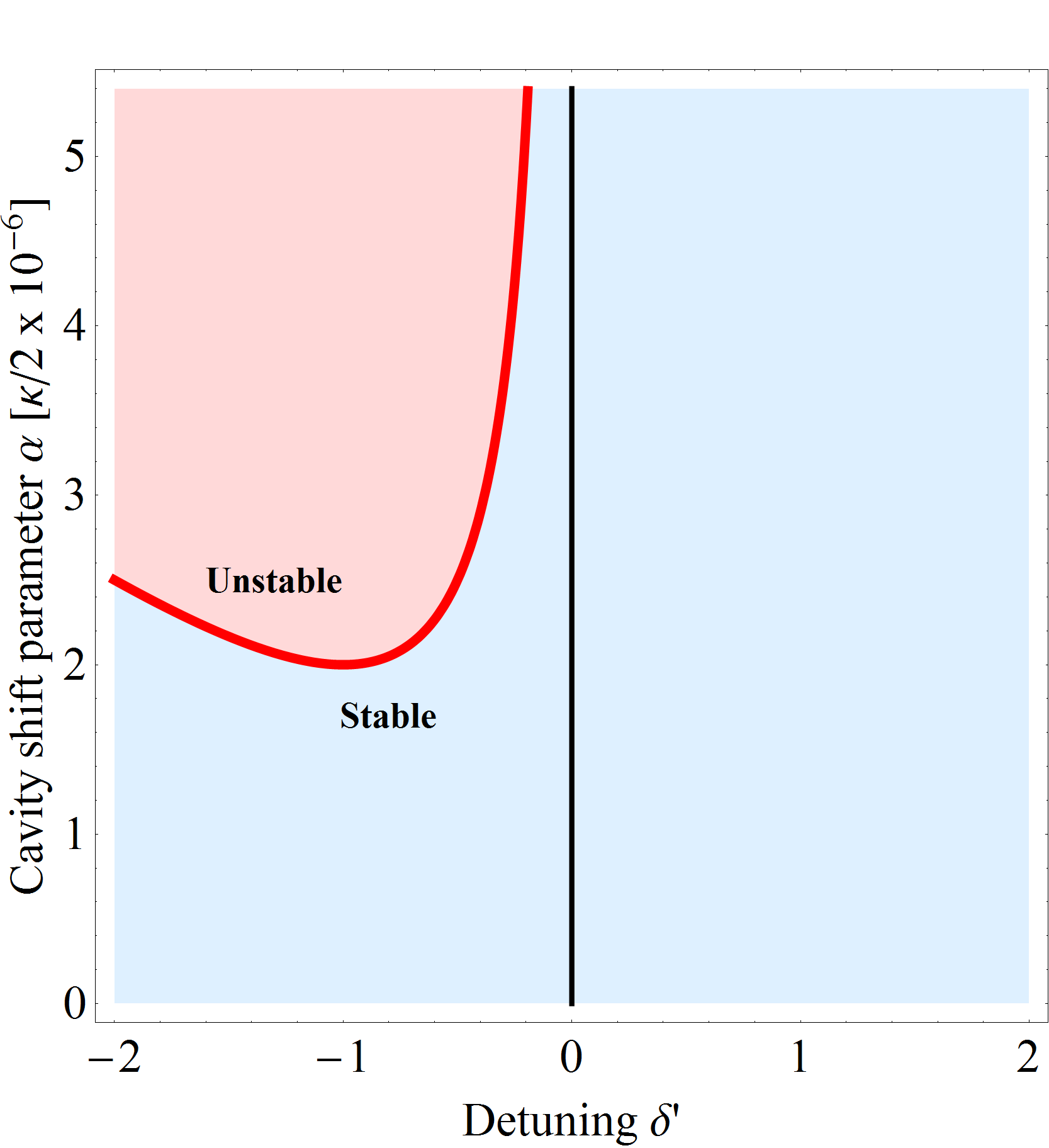}
\caption{Stability plot using $\gamma_0$ stability condition of Eqn. \ref{eqn:gamma0stability} as a function of detuning $\delta'$ and the cavity shift parameter $\alpha$, assuming $N = 10^6$. The stability condition also assumes $W = W_{\mathrm{opt}}(\delta', \Gamma_R)$.  The region of stability is exact for the two-level model ($r=\infty$), and a good approximation for all values of $r$. The blue region shows where the real part of all the poles of the $\jmath_\perp$ solution are negative, indicating a damped return to steady-state conditions for a perturbation. The red region shows where any of the real parts of the poles become positive, making $J_\perp$ unstable, with no steady-state solutions.}
\label{fig:TwoLevelStability}
\end{figure}

\begin{figure}
\includegraphics[width=3in]{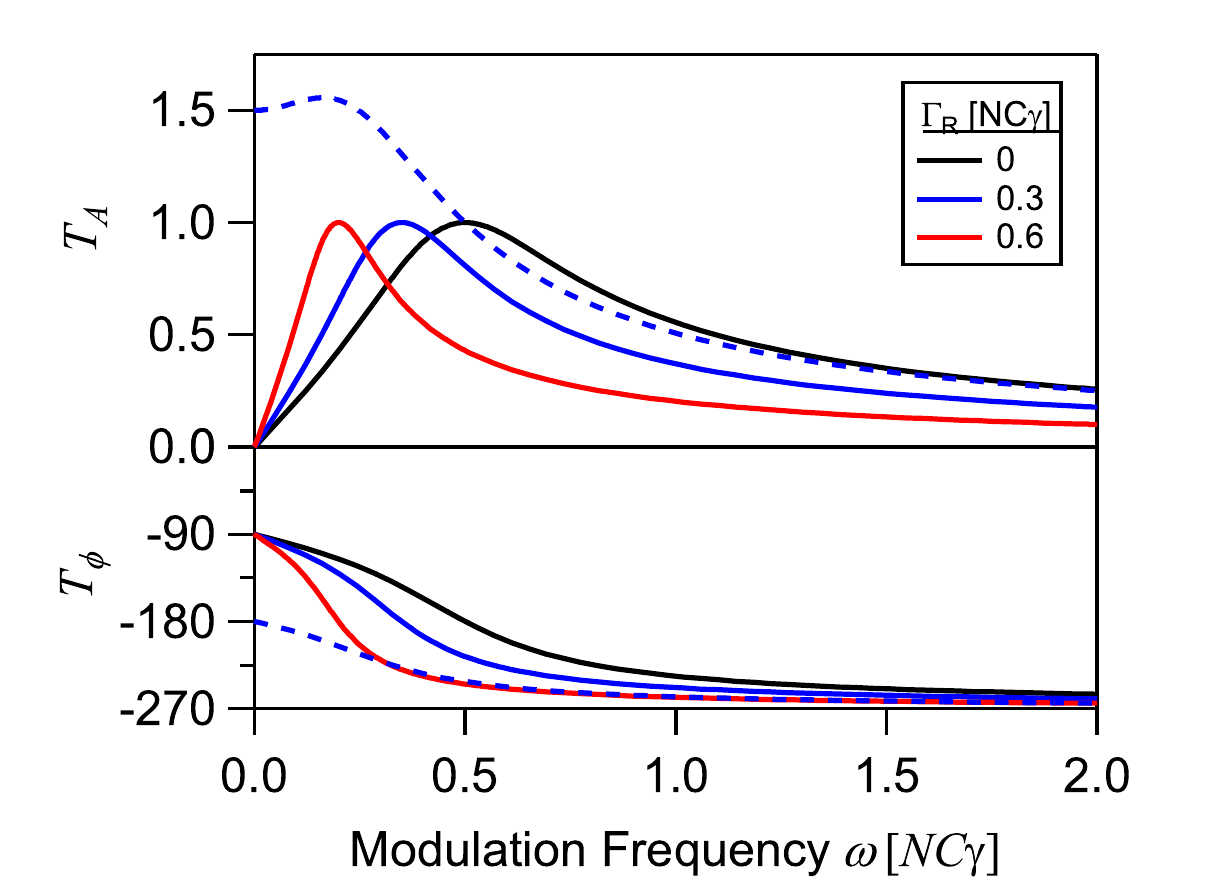}
\caption{Output photon flux transfer function for different Rayleigh scattering rates $\Gamma_R$, with $\xbar{W} = W_{opt}(\Gamma_R)$, $r = \infty$, $\alpha = 0$ and $\delta' = 0$.  The dashed line shows the transfer functions when $\xbar{W}$ is held to $NC\gamma/2$, not varied to remain at $W_{opt}$, and $\Gamma_R = 0.3$.  A dashed red curve is not shown, as with $\Gamma_R = 0.6$ and $\xbar{W} = NC\gamma/2$ the maximum repumping rate threshold has been exceeded and the output photon flux is zero. }
\label{fig:TfuncVsGR}
\end{figure}

We analyze the response of the cavity field amplitude to an applied modulation of the repumping rates by plotting the amplitude transfer function and the phase transfer function versus the modulation frequency $\omega$, defined  as $T_A(\omega) \equiv |a|/\epsilon$ and $T_\phi(\omega) \equiv \arctan\left(\frac{Re[a]}{Im[a]}\right)$ respectively.  We consider the maximum of the transfer function to define the resonant frequency $\omega_{\mathrm{res}}$.  The calculated variation in the transfer functions versus various experimental parameters is shown in Figs. \ref{fig:TfuncVsW}~-~\ref{fig:TfuncVsGR}.  All results are given as a series of transfer functions varying a single specified system parameter, with other unspecified parameters set to $\xbar{W} = W_{\mathrm{opt}}$, $r=\infty$, $\delta' = 0$, and $\Gamma_R = 0$.

The expressions for the damping $\gamma_0$ and the natural frequency $\omega_0$ guide our understanding of the transfer functions. Holding $r=\infty$, $\delta' = 0$, and $\Gamma_R = 0$, the damping reduces to $\gamma_0 =\xbar{W}/2$. Physically, the damping enters through the decay of $J_\perp$ at a rate proportional to $\gamma_\perp$. The natural frequency  $\omega_0 = \sqrt{\xbar{W}(NC\gamma-\xbar{W})} = \sqrt{2}\bar{J}_\perp C\gamma$ is set by the steady-state rate of converting collective transverse coherence into atoms in the ground state, $J_\perp^2C'\gamma$, normalized by the steady-state transverse coherence $J_\perp$.

To examine the effect of the steady-state repumping rate $\xbar{W}$ on the response, we plot the transfer functions $T_A$ and $T_\phi$ for different values of $\xbar{W}$ in Fig. \ref{fig:TfuncVsW}.  For $\xbar{W} < W_{opt}$, we see a narrow resonance feature in the response(blue curve).  The frequency of the resonance increases until $\xbar{W} = W_{opt}$ (green curve). Also at $\xbar{W} = W_{opt}$, the dc amplitude response $T_A(\omega = 0) =0$, because the drive $D_\perp$ goes to zero (Eqn. \ref{eqn:D_perpSimple}), consistent with the maximum in $\dot{M}_c$ at $W_{opt}$. For $\xbar{W} > W_{opt}$, the phase of the response near dc sharply changes sign, as understood from the parabolic response of $\dot{M}_c$ versus $\xbar{W}$; on the $\xbar{W}>W_{opt}$ side of the parabola, the same change in $W$ produces the opposite change in the output photon flux compared to the $\xbar{W}<W_{opt}$ side of the parabola.  Meanwhile, the natural frequency has decreased with the increase in $W$ when $W>W_{opt}$.  As $\xbar{W}$ approaches $W_{\mathrm{max}}$, the response has essentially become that of a single-pole, low pass filter with an additional $\pi$ phase shift.  

To examine the effect of population in the third state $\ket{3}$, we now hold $\xbar{W} = W_{opt}$ and show $T_A$ and $T_\phi$ for different $r$ in Fig. \ref{fig:TfuncVsr}.  The black curve shows the result for $r=\infty$, which is the two-level model of Ref. \cite{MYC09}, as no population accumulates in \ket{3} (recall that $\bar{N}_3/\bar{N}_g =1/r$).  For smaller $r$, the relaxation oscillations grow, shown by the increasing maximum in $T_A$. This response is consistent with the reduced damping rate $\gamma_0$ and increased drive $D_\perp$ seen in the following expressions. 

The damping is $\gamma_0 = \frac{r}{1+r}\left(\frac{NC\gamma}{4}\right)-\frac{2\omega^2}{NC\gamma(1+r)}$.  The additional $\omega$ dependence, associated with the repumping delay from atoms spending time in \ket{3}, results from the third derivative term that scales with $\beta = \frac{1}{\xbar{W}(1+r)}$ in Eqns. \ref{eqn:DiffEqResponse1} and \ref{eqn:DiffEqResponse2}.

The complex drive in this limit is $D_\perp = i\omega \frac{NC\gamma}{2} \frac{1+r+2r^2}{(1+r)(1+2r)} - \frac{\omega^2}{1+r}$.  The term proportional to $\omega^2$ in $D_\perp$ arises from modulating the rate out of the state \ket{3}. Although the $\omega^2$ term in the damping would introduce a roll off in the transfer function $T_A$ with the form $1/\omega^2$, the $\omega^2$ frequency dependence is canceled. The final transfer function maintains a frequency dependence of $1/\omega$ for $\omega \gg \omega_{res}$, similar to that of the two-level system.

Next we consider the effect of the dynamically tunable cavity mode. The cavity mode response can strongly modify the damping of the oscillator and even lead to instabilities in the cavity light field, eliminating steady-state solutions.  We first consider the damping rate of the two-level model ($r = \infty$) with cavity tuning, $\gamma_0 =  \frac{\xbar{W}}{2} \left(1+h(\delta')\right)$ where $h(\delta') = 2\alpha\delta'\left(\frac{N}{1+\delta'^2} -\frac{\xbar{W}+\Gamma_R}{C\gamma}\right)$. The damping is modified by a detuning dependent feedback factor $h(\delta')$ that is positive or negative depending on the sign of $\delta'$.  Because $\xbar{W}+\Gamma_R < \frac{NC\gamma}{1+\delta'^2}$ to meet superradiant threshold, $h(\delta')$ has the same sign as $\delta'$.  Applying negative cavity feedback, when $h(\delta')>0$, increases the damping and may be useful for reducing relaxation oscillations and suppressing the effect of external perturbations. When $h(\delta')<0$, positive feedback decreases $\gamma_0$ and amplifies the effect of perturbations. 

We show the effect of this cavity feedback on the transfer functions in Fig. \ref{fig:TfuncVsd} for the conditions $r=\infty$, $\xbar{W} = W_{opt}(\delta')$, and $\Gamma_R = 0$. The red (blue) curves show positive (negative) feedback, with the black curve serving again as a reference to the model of Ref. \cite{MYC09} with no cavity feedback.

Fig. \ref{fig:TfuncVsd} also shows the effect of increasing the cavity shift parameter $\alpha$. The solid lines result from $\alpha = NC\gamma \times 10^{-3}$, a cavity shift similar in magnitude to experiments performed in Refs. \cite{BCW12,BCWDyn,BCWHybrid,WCB12}. The dashed lines result when $\alpha$ is increased by a factor of two.

With enough positive feedback, the system can become unstable, with any perturbations exponentially growing instead of damping, which eliminates steady-state solutions. For a driven harmonic oscillator, the condition for steady-state solutions is $\gamma_0 > 0$. Again assuming $W = W_{\mathrm{opt}}(\Gamma_R,\delta') = \frac{NC\gamma}{2(1+\delta'^2)}-\Gamma_R$, and remaining in the two level limit ($r=\infty$) the stability condition reduces to

\begin{equation}
N\frac{\alpha \delta'}{1+\delta'^2} > -1 \,\, .
\label{eqn:gamma0stability}
\end{equation}

\noindent In Fig. \ref{fig:TwoLevelStability}, we plot the stability condition as a red line. 

In general, the stability of a linear system can be determined by examining the poles of the solution. If any pole crosses into the right half of the complex plane, the system is unstable with an oscillating solution that grows exponentially. In the two-level limit ($r=\infty$), this condition on the solutions $\jmath_\perp$ and $\jmath_z$ is mathematically equivalent to the condition on $\gamma_0$, Eqn. \ref{eqn:gamma0stability}.  As the level structure becomes more complex, e.g. $r \not= \infty$ or in the full $^{87}$Rb model in Sec. IV, we use the pole analysis to examine the regions of stable operation.  For the model here, as $r$ changes, the pole analysis shows that the stability condition in Eqn. \ref{eqn:gamma0stability} is no longer exactly correct. However, the change is small enough that Eqn. \ref{eqn:gamma0stability} remains a good approximation of the stability condition for all values of $r$.

\begin{figure*}
\includegraphics[width=5.75in]{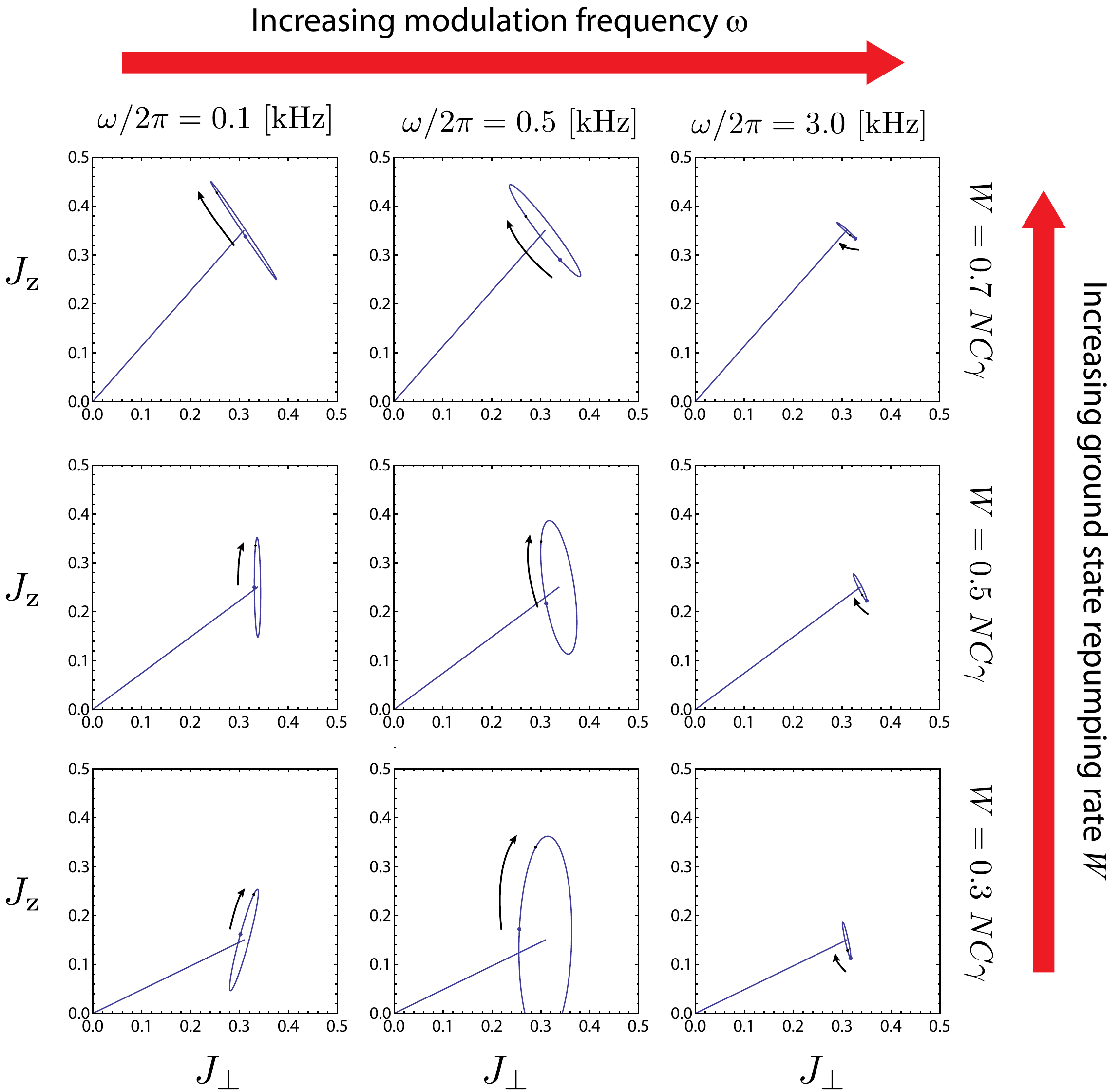}
\caption{Response of the 2D Bloch vector to external modulation of the repumping rate.  The steady-state Bloch vector, i.e. $\bar{J}_\perp$ and $\bar{J}_z$ from Eqns. \ref{eqn:SSJz}-\ref{eqn:SSJperp}, is indicated by the blue line, plotted on the axis with units of $N$, so $N/2$ is the maximum value. The ellipse is the trajectory of the Bloch vector responding to the modulation of the repumping rate $w(t) = \epsilon \mathrm{Re}[e^{i\omega t}]$, described by the small signal responses $\jmath_\perp$ and $\jmath_z$ in Eqns. \ref{eqn:DiffEqResponse1} and \ref{eqn:DiffEqResponse2}. The parameters are $\epsilon = 0.1$, $r=5$, $\delta' = 0$, $\alpha = 0$, and $\Gamma_R = 0$. The black arrow indicates the direction of the trajectory, starting from the blue dot at $t = 0$. The values of $\omega$ are chosen to show $\omega \ll \omega_{\mathrm{res}}$, $\omega \approx \omega_{\mathrm{res}}$, and $\omega \gg \omega_{\mathrm{res}}$.
 }
\label{fig:BlochVectors}
\end{figure*}

\begin{figure*}
\includegraphics[width=7.1in]{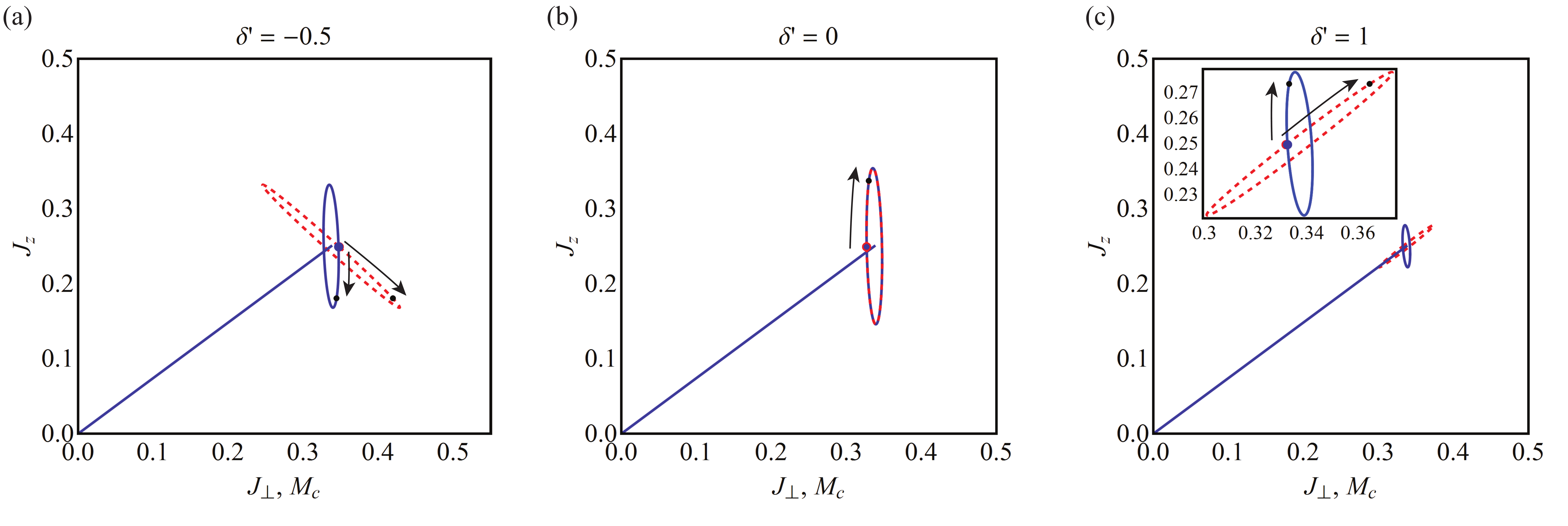}
\caption{Parametric plots of response of the three degrees of freedom $J_z$, $J_\perp$, and $A$, highlighting effect of cavity frequency tuning on response of atomic coherence and output light field. The blue line represents the steady-state atomic Bloch vector, $\bar{J}_z$ and $\bar{J}_\perp$, from Eqns. \ref{eqn:SSJz}-\ref{eqn:SSJperp}.  The blue ellipse shows small single response of the Bloch vector to a modulation of the repumping rate $W$, given by $\jmath_z$ and $\jmath_\perp$ from Eqns. \ref{eqn:DiffEqResponse1} and \ref{eqn:DiffEqResponse2}. The parametric response is plotted with units of $N$, so $N/2$ is the maximum value. The red dashed line is the trajectory formed by the response of the cavity field $a$ (Eqn. \ref{eqn:cavityfieldresponse}) and $\jmath_z$. The cavity field is plotted as a fraction of the average field, then centered on the steady-state Bloch vector to compare with the atomic response.  The arrows indicate the direction of the response with respect to a modulation $W(t) = \xbar{W}(1+\epsilon e^{i \omega t})$.  Here $NC\gamma = 10^4$ s$^{-1}$, $r = 5$, $\Gamma_R = 0$, $\xbar{W} = W_{\mathrm{opt}}(\delta')$, $\epsilon = 0.1$, and $\omega = 0.02\,NC\gamma$, chosen to show the stable $J_\perp$ response. (a) When $\delta' < 0$, the cavity feedback can be positive, leading to larger oscillations compared to the case of no feedback $\delta' =0$ shown in (b).  Because of the coupling of $J_z$ to the cavity mode frequency, $A$ is not locked to the $J_\perp$ response, as in (b), but is anti-correlated with $J_z$.  In (c), where $\delta' > 0$, the negative feedback reduces the response amplitudes in all quadratures.  The cavity tuning again shifts the cavity amplitude response, but with the opposite phase relationship due the change in sign of the slope of the Lorentzian, so $A$ follows $J_z$. The inset shows a close up of the response.}
\label{fig:detuningphasor}
\end{figure*}

Finally,  in Fig. \ref{fig:TfuncVsGR} we show the effect of additional decoherence by plotting $T_A$ and $T_\phi$ for different values of $\Gamma_R$. Here $r = \infty$, $\delta' = 0$, and $\alpha = 0$.  As a reference, the black curve shows the transfer function with $\Gamma_R = 0$. For the solid curves, the ground state repumping rate $\xbar{W} = W_{opt}(\Gamma_R)$ is varied with $\Gamma_R$ to remain at the point of maximum output power (Fig. \ref{fig:fluxVsGR}) which amounts to holding $\gamma_\perp$ constant. Thus, as the decoherence increases by increasing the rate of Rayleigh scattering from the ground state, the resonance frequency only moves because $\xbar{W}$ is changing, as seen in the expression for the natural frequency $\omega_0 = \W(NC\gamma-\Wp)$.  Notice that additional decoherence does not affect the peak size of the relaxation oscillations.  Although the damping rate decreases because $\gamma_0 = \xbar{W}/2$, this effect is canceled by the drive decreasing with $\xbar{W}$ as well, with $D_\perp =  - i\omega (\xbar{W}/2)$ when $\xbar{W} = W_{opt}$.

If we hold $\xbar{W}$ constant at $NC\gamma/2$, the resulting transfer function is the dashed line in Fig. \ref{fig:TfuncVsGR}.  With $\xbar{W}$ constant, the coherence damping rate $\gamma_\perp$ varies with $\Gamma_R$, and the response actually behaves similar to the case where $\xbar{W}$ is increased (Fig. \ref{fig:TfuncVsW}) because of the symmetric roles $W$ and $\Gamma_R$ have in the natural frequency and the drive.

The main conclusion from our examination of the linear response theory of the three-level, bad cavity laser is that most conditions for optimizing the output power are compatible with an amplitude stable laser.  Operating at the optimum repumping rate in particular suppresses the impact of low frequency noise on the amplitude stability.  However, we also find that because the cavity detuning $\delta'$ couples to the population of the laser levels, cavity feedback can act to suppress perturbations, or cause unstable operation, depending on the sign of $\delta'$. A simple relationship between $N$, $\delta'$, and $\alpha$ gives the condition for stable operation at $\xbar{W} = W_{\mathrm{opt}}$.

\subsection{E. Bloch vector analysis of response}

Relaxation oscillations in a good-cavity laser arise from two coupled degrees of freedom,  the intracavity field $A$ and the atomic inversion $J_z$, responding to perturbations at comparable rates. Parametric plots of the amplitude and inversion response provide more insight into the nature of the relaxation oscillations than looking at the laser field amplitude response alone\cite{siegman86}. In the bad-cavity regime, the cavity-field $A$ adiabatically follows the atomic coherence $J_\perp$, and the oscillations arise from a coupling of $J_\perp$ and the inversion $J_z$.  Thus the relevant parametric plot is the 2D projection of the 3D Bloch vector in the rotating frame of the azimuthal angle. In this section, we study this response of the Bloch vector to better understand the stability of the bad-cavity laser.
 

The individual plots of Fig. \ref{fig:BlochVectors} show the trajectory of the Bloch vector for the small signal response at different applied modulation frequencies $\omega$ and different repumping rates $\xbar{W}$.  The trajectory is calculated using the amplitude and phase quadratures of the responses $\jmath_\perp$ and $\jmath_z$ to define the sinusoidal variation of each quadrature with respect to a sinusoidal modulation of $W(t) = \xbar{W}(1+\epsilon \mathrm{Re}[e^{i \omega t}])$. The series of plots show the trend in the responses versus the ground state repumping rate $\xbar{W}$ and modulation frequency $\omega$, with $r = \infty$, $\Gamma_R = 0$, and $\delta' = 0$.  Although the oscillator characteristics of the two quadratures are identical, they display a differing phase in their response due to the differences in the drives $D_\perp$, $D_z$ on the two quadratures.

At high repumping rates $\xbar{W} > W_{\mathrm{opt}}$ and high modulation frequencies $\omega > \omega_{\mathrm{res}}$, the perturbation modulates the polar angle of the Bloch vector, leaving the length largely unchanged.  Near $\omega_{\mathrm{res}}$, the two quadratures have large amplitudes and oscillate close to 90$^\circ$ out of phase, leading to the trajectories that encloses a large area. When $\omega < \omega_{\mathrm{res}}$ and with $\xbar{W}$ near $W_{opt}$, the cancellation in the drive term $D_\perp$ leads to almost no amplitude of oscillation in the $J_\perp$ quadrature, making the modulation predominately $J_z$-like.  For $\alpha=0$ or $\delta' = 0$, this means the cavity field amplitude $A$ will also be stabilized, as it is locked to the transverse coherence $J_\perp$ (Eqn. \ref{eqn:cavityfieldresponse}).

\begin{figure}
\includegraphics[width=2.75in]{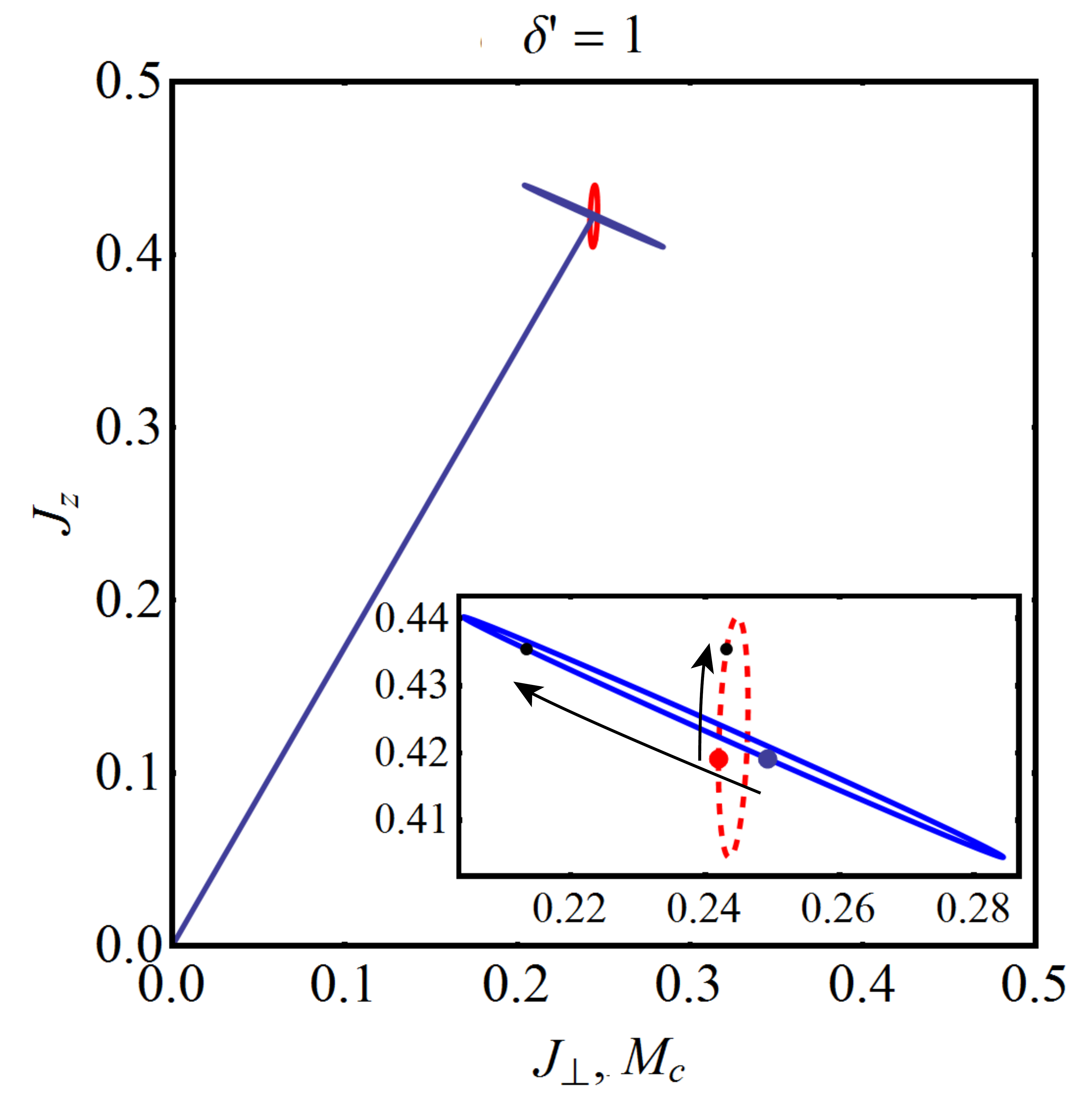}
\caption{Cavity tuning stabilizing the cavity field amplitude. By changing the average repumping rate $\xbar{W}$ from $W_{\mathrm{opt}}$ to $0.44\times 10^4$ for the same parameters as Fig. \ref{fig:detuningphasor}c ($NC\gamma = 10^4$ s$^{-1}$, $r = 5$, $\Gamma_R = 0$, $\epsilon = 0.1$, $\delta' = 1$, and $\omega = 0.02\, NC\gamma$), the response of the Bloch vector (blue ellipse) becomes primarily perpendicular to the steady state Bloch vector (blue line). Under these conditions, the cavity field response $A$ (red dashed ellipse) has the smallest fractional deviation of the three degrees of freedom. The inset shows a close up of the response. 
}
\label{fig:detuningphasorfieldstable}
\end{figure}

However, dynamic cavity tuning creates a coupling of the inversion to the cavity field as well, breaking the simple time-independent proportionality of the cavity field amplitude $A$ and the atomic coherence $J_\perp$, as expected from Eqn. \ref{eqn:cavityfieldresponse}.  Fig. \ref{fig:detuningphasor}a show the case of $\delta' < 0$. Because of the coupling to the inversion, the cavity field response has a larger amplitude than $J_\perp$ response in addition to a phase shift.  It is also nearly $180^\circ$ out of phase with the response of the inversion.
We include the case of $\delta' = 0$ (Fig. \ref{fig:detuningphasor}b) as a reference. The cavity field is locked to the coherence, even for $\alpha \not=0$, due to the second order insensitivity in the cavity coupling. For the case of negative feedback $\delta' > 0$, shown in Fig. \ref{fig:detuningphasor}c, all the response amplitudes are reduced due to the increased damping. Notice that the inversion and cavity field are now responding in phase.   

Because of the coupling between all three degrees of freedom, it is possible to choose parameters that lead to a stabilization of the cavity field.  Operating away from $\Wpk$, the response of the Bloch vector becomes primarily a modulation of the polar angle as the inversion and coherence respond 180$^\circ$ out of phase.  Combined with the cavity tuning, the cavity field is stabilized, as shown in Fig. \ref{fig:detuningphasorfieldstable}, where the parametric plot of $A$ and $J_z$ (dashed red ellipse) shows a response that is primarily $J_z$-like.  The response of the cavity field has the smallest fractional variation among the three degrees of freedom.

To conclude our discussion of linear response theory in the three-level model, we point out that the parametric plot analysis highlights the role that the dispersive cavity frequency tuning plays in amplifying or suppressing perturbations in both the atomic degrees of freedom and the cavity field.  Crucially, frequency stable lasers may need to seek a configuration that suppresses fluctuations in the $J_z$ degree of freedom to minimize the impact of cavity pulling on the frequency of the laser.  We also see that the dispersive tuning breaks the exact proportionality of the cavity field and the transverse atomic coherence, restoring an additional degree of freedom that may be crucial for observing chaotic dynamics in lasers operating deep into the bad-cavity regime \cite{Haken197577}.

\section{III. Raman laser system} 
In the previous section, we presented a model for a three level laser for qualitatively describing the results from recent experiments that use laser cooled $^{87}$Rb as the gain medium \cite{BCW12,BCWDyn, BCWHybrid}. However, the $^{87}$Rb system also relies on a two-photon Raman lasing transition between hyperfine ground states, instead of a single optical transition. To address this difference, here we provide a model that has a two-photon Raman lasing transition, but a simple one-step repumping scheme directly from \ket{g} to \ket{e}.  Then in Sec. IV, we present a full model of the bad-cavity laser in  $^{87}$Rb that has both the two-photon Raman transition and a more complex repumping scheme.



In the first subsection, we derive equations of motion for the expectation values in the Raman model, then explicitly adiabatically eliminate the optically excited intermediate state in the Raman transition.  In the second subsection, will establish the equivalences (and differences) between the Raman and non-Raman models. We will find that the Raman transition is well described as a one-photon transition with a spontaneous decay rate $\gamma$, an effective atom-cavity coupling $g_2$, and with a two-photon cooperativity parameter $C_2$ equal to the original one-photon cooperativity parameter.  The Raman system differs in the appearance of two new phenomena:  differential light shifts between ground states and cavity frequency tuning in response to atomic population changes. The latter effect was inserted by hand in Sec II.  As in Sec. II, we first derive equations without assuming a good-cavity or bad-cavity laser, only specializing to the bad-cavity limit at the end of the section.

\begin{figure}
\includegraphics[width=3in]{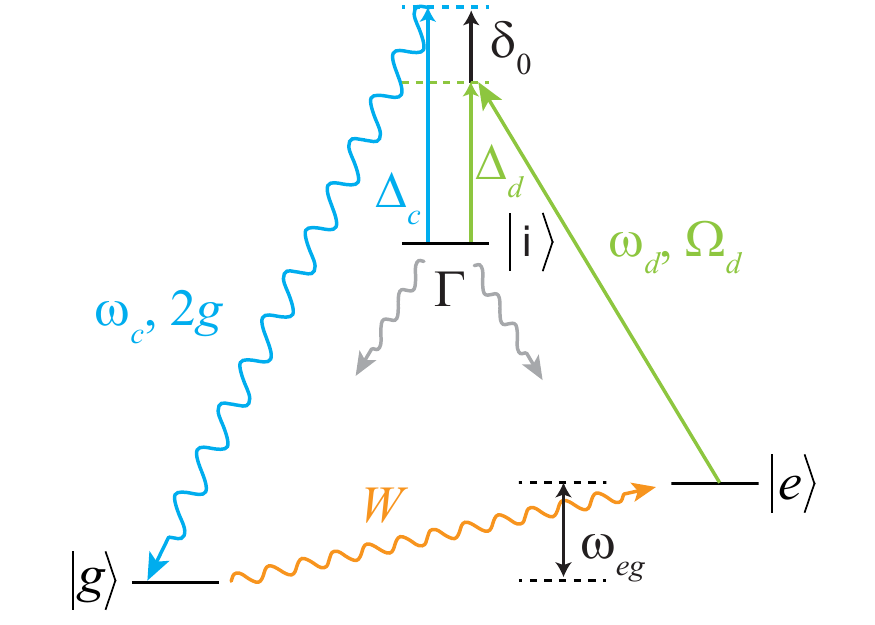}
\caption{Energy level diagram for a superradiant laser enabled by an induced Raman transition.  States \ket{e} and \ket{g} are two metastable states separated by a non-optical frequency $\omega_{eg}$. They share an optically excited state \ket{\mathrm{i}} that has a linewidth $\Gamma$.  Using a Raman dressing laser (green), detuned from \ket{\mathrm{i}} by $\Delta_d$, we can induced a optical decay to \ket{g}, which, in absence of collective effects, would proceed at rate $\gamma = \frac{\Gamma}{4}\left( \frac{\Omega_d}{\Delta}\right)^2$.  Including a single optical cavity mode, coupled to the \ket{\mathrm{i}} to \ket{g} transition with coupling constant $2g$, gives rise to collective emission.  The cavity mode frequency is $\omega_c$, detuned from \ket{\mathrm{i}} by $\Delta_c$, making the two-photon detuning $\delta_0 = \omega_c - (\omega_d + \omega_{eg})$.  To complete the laser cycle, the atoms are incoherently repumped from \ket{g} to \ket{e} at a rate $W$.
}
\label{fig:RamanEnLvl} 
\end{figure}

\subsection{A. Adiabatic elimination of the intermediate state}

To establish the connection between two-photon Raman lasing and one-photon lasing, we start by defining the Hilbert space for a three-level Raman system with two ground states denoted \ket{g} and \ket{e} (separated by only 6.834~GHz in $^{87}$Rb) and an optically excited intermediate state \ket{\mathrm{i}} (Fig. \ref{fig:RamanEnLvl}). The Hilbert space also includes a single cavity mode that couples \ket{g} to \ket{\mathrm{i}}.  The density operator for the Hilbert space is $\hat{\rho}  = \sum_{q=1}^N\sum_{kl}\sum^\infty_{mn}\ket{k^{(q)},n}\bra{l^{(q)},m}$. The first sum is over individual atoms, the second is over the atomic basis states $k,l \in \{i,e,g\}$, and the third sum is over cavity Fock, or photon-number, states.  Raising and lowering operators for the cavity field and atoms are defined as in Sec. II. The state occupation operators for atoms in the state \ket{k} are again $\hat{N}_k = \sum^N_{q=1}\ket{k^{(q)}}\bra{k^{(q)}}$, where the index $q$ denotes a sum over individual atoms. We also define collective atomic raising and lower operators $\hat{J}_{kl} = \sum^N_{q=1} \ket{k^{(q)}}\bra{l^{(q)}}$.


We describe the system via the semi-classical Hamiltonian 

\begin{align}
\begin{split}
H &= \hbar \omega_c \hat{c}^\dag \hat{c} + \hbar \omega_i \hat{N}_i + \hbar \omega_e \hat{N}_e + \hbar \omega_g \hat{N}_g \\ 
&+ \hbar \frac{\Omega_d(t)}{2} (\hat{J}_{ei}+\hat{J}_{ie}) +  \hbar g (\hat{c}^\dag \hat{J}_{gi} + \hat{c}\hat{J}_{ig}).
\label{eqn:Hint}
\end{split}
\end{align}

\noindent The Raman dressing laser at frequency $\omega_d$ is described by the coupling $\Omega_d(t) = \Omega_d(e^{-i \omega_d t}+e^{i \omega_d t})$, and the atoms are uniformly coupled to the dressing laser. The rotating wave approximation will be applied so that only near-resonant interactions will be considered. The dressing field is externally applied, and  we assume it is unaffected by the system dynamics (i.e., there is no depletion of the field).

To reduce the Raman transition to an effective two-level system, we derive the equations of motion for expectation values of the operators that describe the field and the atomic degrees of freedom.  As was done in Sec. II, we use the time evolution of the density matrix obtained from the master equation (Eqn. \ref{eqn:masterEquation}) to derive the equations of motion $\dot{\mathcal{O}} = \Tr[\hat{\mathcal{O}}\hat{\rho}]$. The details are included in Appendix B.

After adiabatic elimination of the optically excited state, we have the set of three coupled equations analogous to Eqns. \ref{eqn:2lvlcavity}-\ref{eqn:2lvldotJz}:

\begin{gather}
\dot{\mathcal{C}} = \left(-\kappa/2 - i \left( \frac{g^2}{\Delta}N_g+\omega_c\right)\right)\mathcal{C}-i \frac{ g \Omega_d}{2\Delta} J_{ge}e^{-i \omega_d t} \label{eqn:labframeCdot}\\ 
\begin{split}
\dot{J}_{ge} &=\left(-\gamma_\perp - i\left(  \frac{\Omega_d^2}{4 \Delta}-\frac{g^2|\mathcal{C}|^2}{\Delta} +\omega_{eg}\right)\right)J_{ge} \\&+ i 2 \frac{g \Omega_d}{2 \Delta} J_z \mathcal{C} e^{i \omega_d t}  \label{eqn:labframeJegdot}
\end{split} \\
\dot{J}_z =W(N/2-J_z) + i \frac{g \Omega_d}{2\Delta}(\mathcal{C}^*J_{ge}e^{-i\omega_d t} - \mathcal{C}J_{eg}e^{i \omega_d t})
\label{eqn:labframeJzdot}
\end{gather}

Here $\gamma_\perp = W/2$ and $\Delta = \Delta_d + (\delta_0/2)$, which is also the average of $\Delta_d$ and $\Delta_c$.  The Eqns. \ref{eqn:labframeCdot}-\ref{eqn:labframeJzdot} are general equations, valid without assuming a good-cavity or bad-cavity laser.

\subsection{B. Defining effective two level parameters for the Raman system}
\noindent We can now identify the effective two-photon atom-cavity coupling constant 

\begin{equation}
g_2 =  \frac{g\Omega_d}{2\Delta}\,\, .
\label{eq:g2}
\end{equation}

\noindent The effective Rabi flopping frequency between \ket{e} and \ket{g} is just $2 g_2$.

Using this coupling constant, we can also construct an effective cooperatively parameter for the two-photon transition using $C_2= (2 g_2)^2/\kappa \gamma$, where 

\begin{equation}
\gamma  = \frac{\Gamma}{4}\left( \frac{\Omega_d}{\Delta}\right)^2
\label{eqn:gamma}
\end{equation}

\noindent is the decay rate for an atom in \ket{e} to \ket{g} induced by the dressing laser, calculated for large detunings.  Substituting Eqns. \ref{eq:g2} and \ref{eqn:gamma} into the above expression for $C_2$, one finds that the two-photon cooperatively parameter and the one-photon cooperatively parameter (Eqn. \ref{eqn:singleparticleCparameter}) are identical $C_2=C=(2 g)^2/\kappa\Gamma$.  This is explained by the geometric interpretation of $C$, a ratio which is determined by the fractional spatial solid angle subtended by the cavity mode and the enhancement provided by the cavity finesse $F$ which enters through the value of $\kappa \propto 1/F$ \cite{TanjiSuzuki2011201}.


The adiabatic elimination yields the two-photon differential ac Stark shift of the frequency difference between \ket{e} and \ket{g}
  
\begin{equation}
\omega_{ac} = \frac{\Omega_d^2}{4\Delta} - \frac{g^2 |\mathcal{C}|^2}{\Delta} \,\, ,
\end{equation}

\noindent seen in Eqn. \ref{eqn:labframeJegdot}.  The two contributions to $\omega_{ac}$ correspond to virtual stimulated absorption and decay. The same virtual process also acts back on the cavity mode creating a cavity frequency as seen in Eqn. \ref{eqn:labframeCdot}.  The shift corresponds to a modification of the bare cavity resonance frequency, leading to a new dressed cavity resonance $\omega_D$ given by

\begin{equation}
\omega_D = \omega_c + N_g\frac{g^2}{\Delta} \,\, .
\label{eqn:dressedCavityFreqRamanOnly}
\end{equation}


\noindent This is the cavity frequency tuning in response to atomic populations artificially introduced in Sec. II.   We have assumed that only an atom in \ket{g} couples to the cavity mode, but in reality both states may couple to the cavity mode such that in general $ \omega_D = \omega_c + N_g\frac{g_g^2}{\Delta_g} + N_e\frac{g_e^2}{\Delta_e}$, where we have specified independent populations, coupling constants, and detunings for the two states \ket{e} and \ket{g} denoted by subscripts.  For tractability in Sec. II's  three-level model,  we assumed the pre-factors $g^2/\Delta$ were equal in magnitude but opposite in sign so that cavity frequency tuning could be written as $\frac{g^2}{\Delta}(N_e-N_g) = \alpha J_z$.

As in Sec. II, we also determine the steady-state frequency of the laser 

\begin{equation}
\omega_\gamma = \frac{2\gamma_\perp}{2\gamma_\perp+\kappa}\omega_D + \frac{\kappa}{2\gamma_\perp+\kappa}(\omega_{eg} + \omega_d -\omega_{ac})\,\, .
\end{equation}

\noindent and define $\delta = \omega_D -\omega_\gamma$ as the detuning of the emission frequency $\omega_\gamma$ from the dressed cavity mode $\omega_D$.


Here we see that, in general, both the atomic transition frequency tuning from $\omega_{ac}$ and the cavity frequency tuning in $\omega_D$ are important for the laser amplitude dynamics.  Comparing the expressions for $\omega_{ac}$ and $\omega_D$, both scale with $g^2/\Delta$, and the determining degrees of freedom are the relative number of atomic to photonic quanta.  In good-cavity systems, a large number of photons can build up in the cavity, and the frequency tuning dynamics are dominated by the ac Stark shift \cite{PhysRevLett.107.063904}.  Superradiant lasers, operating deep in the bad-cavity regime, can operate with less than one intracavity photon on average \cite{BCW12}, resulting in a system with amplitude dynamics dominated by dispersive tuning of the cavity mode from population \cite{BCWDyn}.  Additional energy levels that couple to the dressed cavity mode $\omega_D$ can result in a proliferation in the degrees of freedom for dispersive cavity tuning, resulting in a much richer system than one dominated by ac Stark shifts, which depend only on $|\mathcal{C}|^2$.

To complete the analogy to the non-Raman lasing transitions from the previous section and arrive at equations for the bad-cavity laser dynamics, here we make the bad-cavity approximation $\kappa \gg 2\gamma_\perp$. We again adiabatically eliminate the cavity field amplitude, assuming that it varies slowly compared to the damping rate. We define a normalized detuning $\delta' = \delta/(\kappa/2)$, and use the cooperativity parameter $C$ to describe the coupling. After simplifying, we have a two-level system analogous to Eqns. \ref{eqn:Jz2lvl} and \ref{eqn:Jperp2lvl} in Sec. II

\begin{gather}
\ddt |J_{ge}|^2 =  - 2\gamma_\perp |J_{eg}|^2 + \frac{2 C\gamma}{1+\delta'^2} J_z |J_{ge}|^2\label{eqn:JperpRamanOnly}\\
\ddt J_z = W(N/2-J_z) - \frac{C\gamma}{1+\delta'^2}|J_{ge}|^2 \,\, .
\label{eqn:JzRamanOnly}
\end{gather}

\noindent Note that in the bad-cavity limit, the detuning of the dressed cavity mode from the emission frequency $\delta$ is to good approximation the difference of the dressed cavity resonant frequency and the dressed atomic frequency, modified by a small cavity pulling factor

\begin{equation}
\delta \approx (\omega_D - (\omega_{eg}-\omega_{ac}+\omega_d))\left(1-\frac{W}{\kappa}\right) \, \, .
\end{equation}

Our conclusion is that a Raman superradiant laser can perform as a single-photon superradiant laser with $C_2 = C$, but with a transverse collective coherence that evolves a quantum phase at a frequency set by the separation of the two ground states.  This means that while superradiant Raman lasers based on hyperfine transitions may not be useful for optical frequency references, their tunability and control make them excellent physical ``test-bed" systems for studying cold atom lasers \cite{PhysRevLett.107.063904,BCW12,BCWDyn}.  In addition, the switchable excited state lifetime in a Raman system introduces the possibility of dynamic control in the superradiant emission, useful for novel atomic sensors \cite{BCWHybrid,WCB12}.

\section{IV. Full model in $^{87}$Rb} 

\begin{figure}
\includegraphics[width=2.4in]{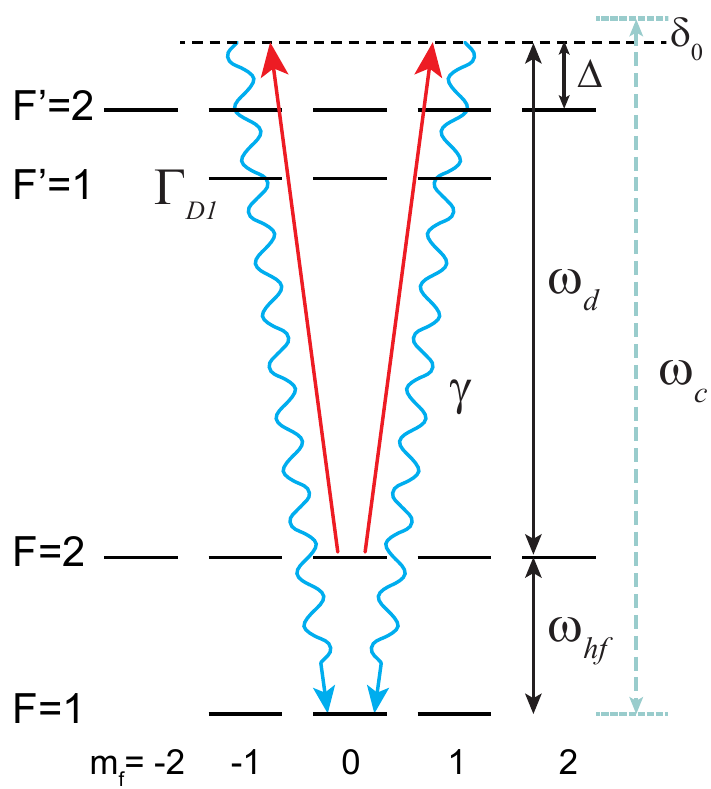}
\caption{Lasing transition and Raman dressing scheme on the $^{87}$Rb D1 line (795 nm). The dressing light (red) and collective emission (blue) are a superposition of $\sigma^+$ and $\sigma^-$ polarizations because the direction of propagation of the light is along the quantization axis defined by the direction of the magnetic field at the atoms. The Raman dressing laser is detuned by $\Delta$ from the atomic transition. The bare cavity detuning is $\delta_0 =\omega_c - ( \omega_d + \omega_{hf})$. The optically excited state on the D1 line has a linewidth $\Gamma_{D1}/2\pi = 5.75$ MHz. The effective population decay from \ket{F=2, m_f=0} to \ket{F=1, m_f=0} is $\gamma = \frac{\Gamma_{D1}}{4}\left( \frac{\Omega_d}{\Delta}\right)^2$.}
\label{RbEnergyLevelLasing}
\end{figure}

\begin{figure}
\includegraphics[width=2.8in]{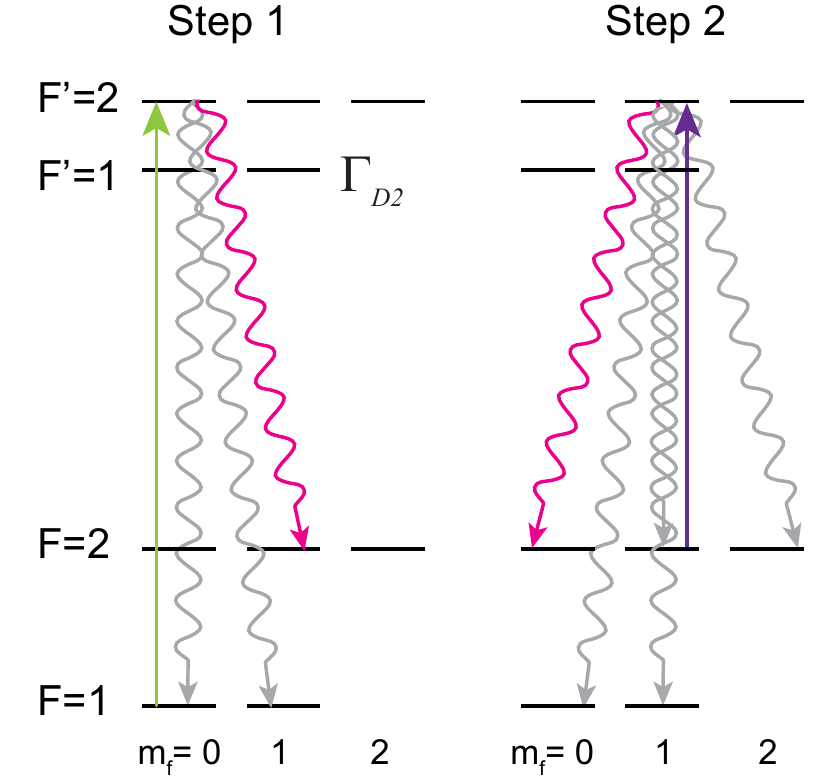}
\caption{Two step repumping process on the $^{87}$Rb D2 line (780 nm). The diagram is drawn showing on only positive Zeeman states, but the process is symmetric with respect to $m_f = \pm 1, \pm 2$. The desirable decay branches (magenta) show the most direct repumping sequence, although any particular repumping sequence could go through many ground hyperfine states. The optically excited state on the D2 line has a linewidth $\Gamma_{D2}/2\pi = 6.07$ MHz.}
\label{RbEnergyLevelRepump}
\end{figure}

In this section, we give the results of a model for a superradiant Raman laser using the ground state hyperfine clock transition (\ket{g} = \ket{5^2S_{1/2}, F=1,m_f=0}, \ket{e} = \ket{5^2S_{1/2}, F=2, m_f=0}) in $^{87}$Rb, including all eight ground state levels for repumping. The results here support the experimental work of Refs. \cite{BCW12, BCWDyn, BCWHybrid, WCB12}, and include specific values of parameters taken from those experiments. The model combines the three-level repumping from the model in Sec. II and the Raman transition between \ket{e} and \ket{g} of Sec. III. After summarizing the key steady-state results, we use linear response theory similar to Sec. II to examine the stability of the laser, identifying the important parameters for stable operation in superradiant Raman lasers.

\subsection{A. Continuous superradiant Raman laser in $^{87}$Rb} 

We model steady-state superradiance  in the full $^{87}$Rb Raman system by first including incoherent repumping among the eight ground $5\,^2S_{1/2}$  hyperfine populations $N_{F, m_F}$ (Fig. \ref{RbEnergyLevelRepump}). We use $\lambda =\{F, m_F\}$ to refer to a generic set of population quantum labels as $N_\lambda$.

The repumping is performed using single particle scattering off optically excited states to result in Raman transitions to move population from $F = 1$ to $F=2$ (Fig. \ref{RbEnergyLevelRepump}). The repumping has a clear analogy to the three-level model  from Sec. II because population cannot be directly transfered from \ket{e} to \ket{g}, meaning some finite population accumulates outside the coherent lasing levels. Separate lasers repump atoms in the $F=1$ state (green) and the $F=2$ state (purple). The lasers are characterized by Rabi frequencies $\Omega_{1,0,2,0}$ and $\Omega_{2,1,2,1}$ respectively. The rate of population transfer out of \ket{g} is proportional to the total scattering rate $W$, which includes the Rayleigh scattering rate. The transverse decoherence rate $\gamma_\perp = W/2$ is dominated by the necessary scattering from repumping. In analogy with the model in Sec. II, the repumping rates out of the states in $F=2$ are proportional to $rW$, where $r = \Omega_{2,1,2,1}^2/\Omega_{1,0,2,0}^2$.  The detailed equations for the repumping are given in Appendix C.

To include the collective emission in our $^{87}$Rb Raman laser model, we reduce the Raman transition dynamics to an effective two-level model by eliminating the optical intermediate state (see Sec. III). The hyperfine ground states \ket{g} and \ket{e} form the effective two-level transition shown in Fig. \ref{RbEnergyLevelLasing}.  The optical transition is induced by a $795$ nm dressing laser with Rabi frequency $\Omega_d$ far detuned from the $\ket{e} \rightarrow \ket{5^2P_{3/2},F'=2, m_F' }$ transition ($\Delta$ is typically 1-2 GHz).  The dressing laser creates an effective spontaneous scattering rate from \ket{e} to \ket{g} $\gamma = \frac{\Gamma_{D1}}{4}\left( \frac{\Omega_d}{\Delta}\right)^2$ (see Eqn. \ref{eqn:gamma} in Sec. III). The fraction of this single particle scattering that goes into the cavity mode is given by the cooperativity parameter $C = 8 \times 10^{-3}$.

Single particle scattering in the cavity mode results in a build up of collective coherence $J_\perp^2$ between \ket{e} and \ket{g}. The collective emission has an enhanced scattering rate which dominates the population transfer from \ket{e} to \ket{g}. We include the population transfer from collective emission along with the equation for the collective coherence Eqn. \ref{eqn:JperpRamanOnly} with the population equations from repumping to form the set of equations used to obtain the steady-state solutions and perform the linearized analysis. We give the details in Appendix C.

\subsection{B. Steady-state solutions}
In analogy to the model in Sec. II, we are concerned with steady-state values of the inversion $J_z = \frac{1}{2}(N_{2,0} - N_{1,0})$, the collective transverse coherence $J_\perp^2$, and the population that occupies energy levels outside the laser transition $N_{other} = N - N_{2,0} - N_{1,0}$. The steady-state solutions of the system equations are

\begin{equation}
\bar{J}^2_{\perp} = \left(\frac{N}{2}\right)^2 \left(\frac{\frac{3}{13}r}{\frac{27}{104}+r}\right) \left(\frac{2W(1+\delta'^2)}{NC\gamma}\right) \left(1-\frac{W(1+\delta'^2)}{NC\gamma}\right)
\label{JpSS}
\end{equation}

\begin{equation}
\bar{J}_{z} = \frac{W(1+\delta'^2)}{2C\gamma}
\end{equation}

\begin{equation}
\bar{N}_{other} = N\left(\frac{4}{13}\right)\left(\frac{\frac{27}{32}+r}{\frac{27}{104}+r}\right) \left(1-\frac{W(1+\delta'^2)}{NC\gamma}\right) 
\label{eqn:Nother}
\end{equation}

\begin{equation}
\bar{M}_c = \bar{J}_\perp^2 \frac{C \gamma}{1+\delta'^2}\,\, .
\end{equation}

\noindent Here $\delta'$ is the detuning of the dressed cavity mode from the laser emission frequency.

As in Sec. II, there is again both a repumping rate that maximizes the coherence (along with the output photon flux) and a repumping threshold for laser turnoff 

\begin{gather}
W_{opt} = \frac{NC\gamma}{2(1+\delta'^2)},\\ 
W_{max} = 2W_{opt}.
\end{gather}

To understand the effect of repumping in the full $^{87}$Rb model, we compare Eqn. \ref{JpSS} to the steady-state coherence in the three-level model, Eqn. \ref{eqn:SSJperp} in Sec. II. While the form of the expression versus the ground state repumping rate $W$ is the same as the three-level model, the scale factor associated with the repumping ratio is modified.  The power reduction factor 

\begin{equation}
R_{Rb}(r) \equiv \frac{3}{13}\left(\frac{r}{\frac{27}{104}+r}\right)
\label{eqn:Rfactor}
\end{equation}

\noindent is the modification to the steady-state photon flux compared to the ideal model in Refs. \cite{MYC09}. $R_{Rb}$ has a maximum value of $3/13$ contrasted with $R$, Eqn. \ref{eqn:Rfactor3lvl}, which has a maximum of 1.
 
While the repuming ratio $r \not\ll 1$, most of the population remains in $N_{2,0}$ and $N_{1,0}$ (Eqn. \ref{eqn:Nother}).  The inversion $J_z$ is the same as the model from Sec. II (Eqn. \ref{eqn:SSJz}), as here $W$ corresponds to $2\gamma_\perp$. Thus, the results of Figs. \ref{fig:fluxVsr}-\ref{fig:fluxVsGR} give a good qualitative understanding of the steady-state behavior of the $^{87}$Rb system as well. 

\begin{figure}
\includegraphics[width=3in]{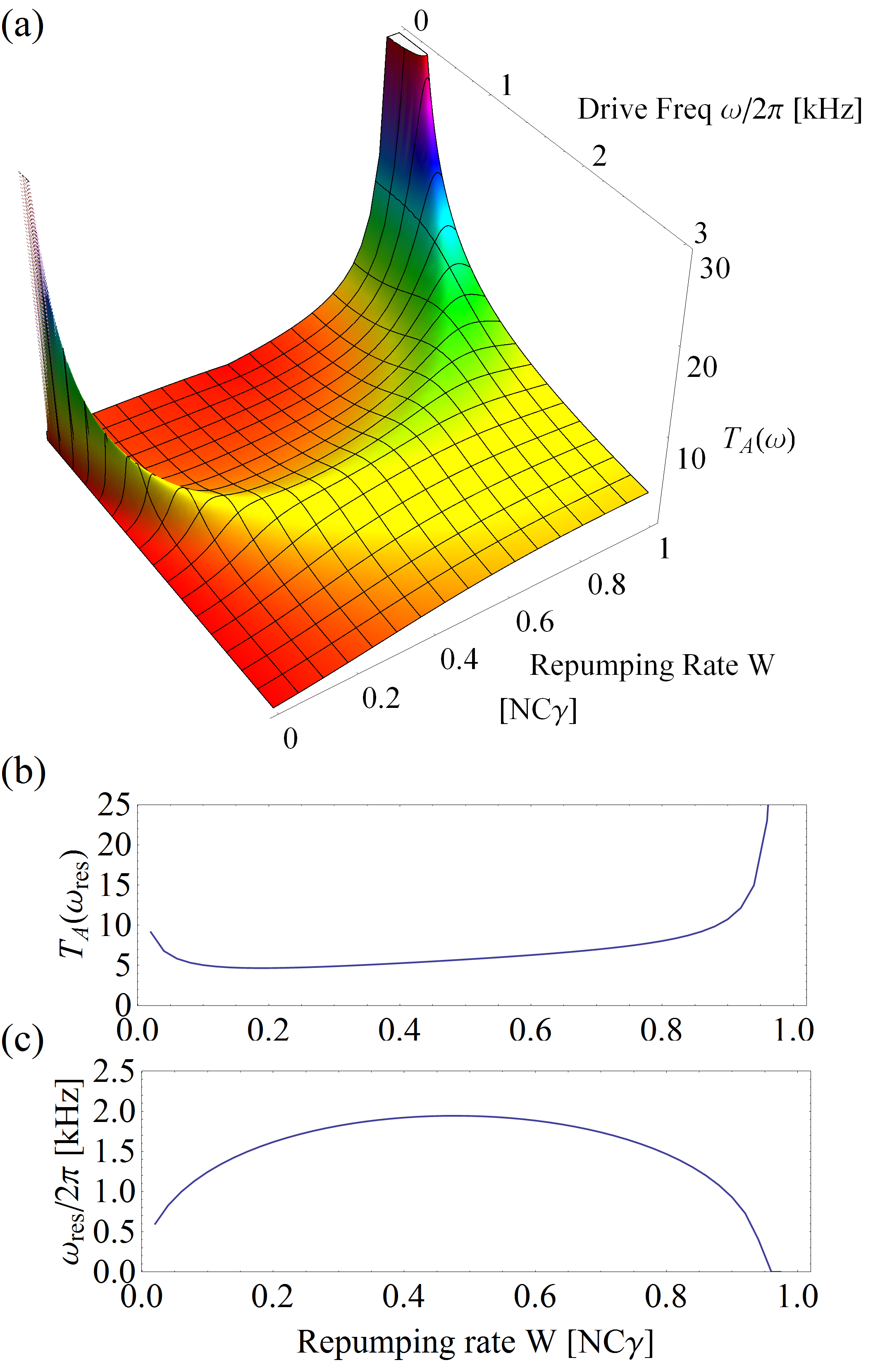}
\caption{Response for different repumping rates. (a) Light field amplitude response transfer function $T_A$ versus repumping rate $W$. (b) Resonant response $T_A(\omega_{max})$ and (c) the resonant modulation frequency $\omega_{max}$ as a function of repumping rate $W$.  Here $NC\gamma = 4\times10^5$ s$^{-1}$, $\delta' = \alpha = 0$, and $r = 0.71$.}
\label{fig:vsW}
\end{figure}

\begin{figure}[b!]
\includegraphics[width=3in]{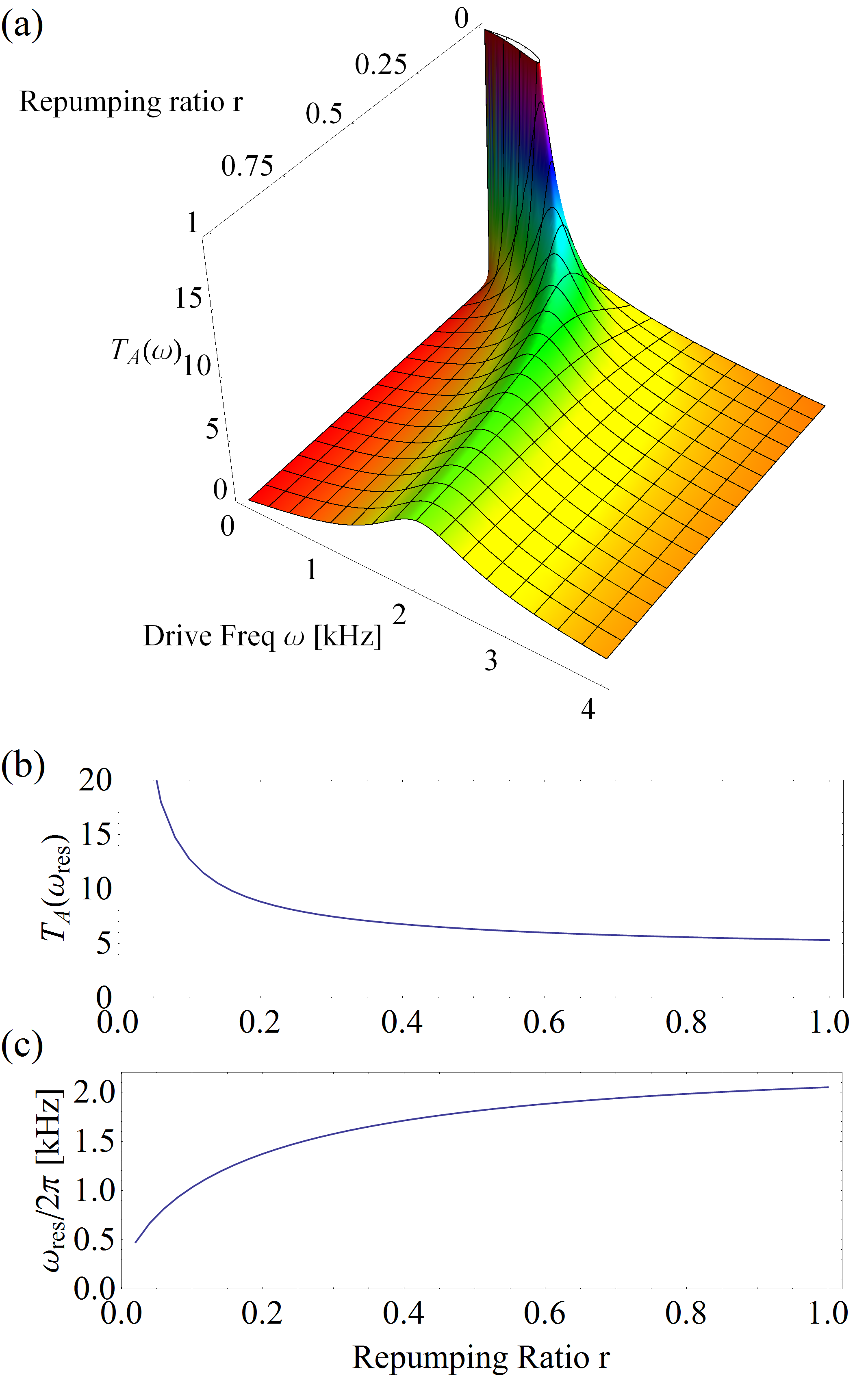}
\caption{Response for different repumping ratio. (a) Light field amplitude response transfer function $T_A$ versus repumping ratio $r$. (b) Resonant response $T_A(\omega_{max})$ and (c) the resonant modulation frequency $\omega_{max}$ as a function of repumping ratio $r$. Here $NC\gamma = 4\times10^5$ s$^{-1}$, $\delta' = 0$ and $\xbar{W} = W_{opt}$.}
\label{fig:vsr}
\end{figure}

\begin{figure}
\includegraphics[width=3.0in]{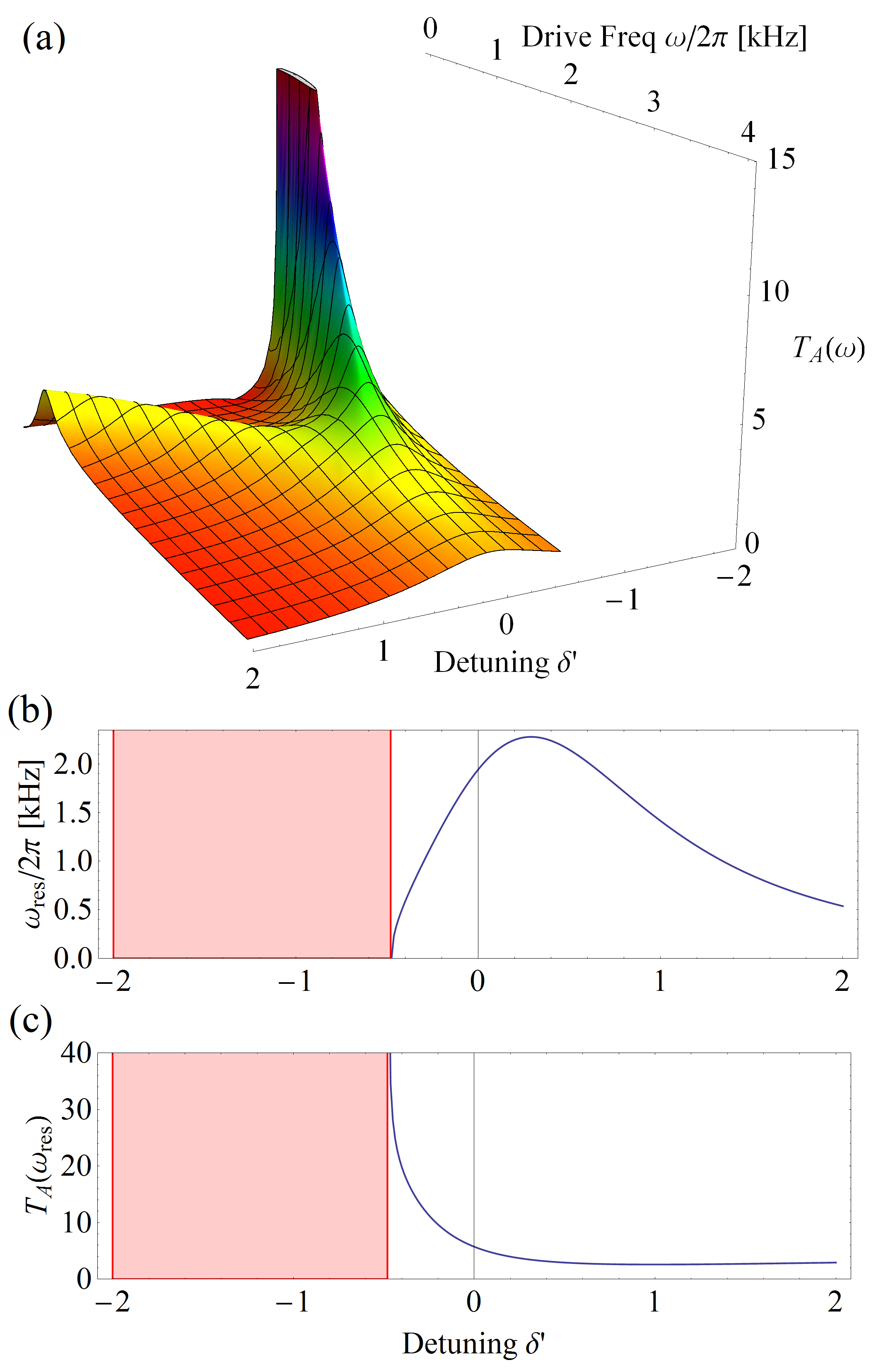}
\caption{Response for different detunings. (a) Light field amplitude response transfer function $T_A$ versus detuning $\delta'$. The transfer function is not plotted in regions of instability. (b) Resonant response $T_A(\omega_{max})$ and (c) the resonant modulation frequency $\omega_{max}$ as a function of $\delta'$. The red shaded regions indicate parameters in which the system is unstable and no steady-state solutions exist. Here $NC\gamma = 4\times10^5$ s$^{-1}$, $r= 0.71$ and $\xbar{W} =W_{opt}$. The cavity shift parameter is given by $\vec{\alpha}_+$. These parameter values reflect the conditions of the experimental system in Ref. \cite{BCWDyn}.}
\label{fig:vsDelta}
\end{figure}

\subsection{C. Linear response theory in $^{87}$Rb}

To analyze the small signal response about these steady-state solutions analytically, we perform the analogous linear expansion as was done in Sec. II. We assume the repumping rates are modulated with $W(t)~=~\xbar{W}(1+\epsilon \mathrm{Re}[e^{i\omega t}])$, and assume the resulting modulation of the populations and coherence take the form $N_\lambda = \bar{N}_{\lambda}(1+n_\lambda(t))$ and $\Jperp = \bar{J}^2_{\perp}(1+\jmath_\perp(t))$.  The equations are then linearized by expanding to first order in the small quantities $n_\lambda(t)$, $\jmath_\perp(t)$, and $\epsilon$, and then re-expressed in terms of $\jmath_z(t) = (\bar{N}_{e} n_e(t) -\bar{N}_{g} n_g(t) )/(\bar{N}_{e}-\bar{N}_{g})$ and $\jmath_\perp(t)$.

We solve for the steady-state, complex response amplitude at a single drive frequency $\jmath_z(t) = \jmath_z(\omega) e^{i\omega t}$,  $\jmath_\perp(t) = \jmath_\perp(\omega) e^{i\omega t} \approx \jmath^2_\perp(\omega) e^{i\omega t}/2$ and $n_\lambda(t) = n_\lambda(\omega) e^{i \omega t}$. The response of the photon amplitude flux is $a(\omega) = \jmath_\perp(\omega) - \frac{\delta'}{1+\delta'} d\delta(\omega)$, where the detuning response is defined by the population response $d\delta(\omega) = \sum_\lambda (\alpha_\lambda/\kappa)N_\lambda n_\lambda(\omega)$ and the $\alpha_\lambda$ are given by elements of the cavity tuning vector  $\vec{\alpha}_+$, given in Appendix C, section II. The predicted normalized fractional amplitude response is $T_A(\omega) = |a|/\epsilon$ and phase response function is $T_\phi(\omega) = \arctan\left( \frac{\text{Re}[a]}{\text{Im}[a]}\right)$.

Figs. \ref{fig:vsW} - \ref{fig:vsDelta} contain surface plots showing the light amplitude transfer function $T_A$ versus modulation frequency.  The third dimension shows how the response changes when a single parameter $\xbar{W}$, $r$, and $\delta'$ is varied.  The lower plots in each figure show the resonant response of the system, following the frequency of the maximum response $\omega_{res}$ and the resonant amplitude response $T_A(\omega_{res})$. The response functions follow the same general trends as the three-level model in Sec. II, showing that the simplified model captures the essential physics of our system.  The full model also demonstrates good quantitative agreement with the experimental results as shown in Ref. \cite{BCWDyn}. 

In Fig. \ref{fig:vsW}, we show the amplitude transfer function versus the repumping rate $\xbar{W}$ assuming $r = 0.71$ and $\delta' = 0$.  The value of $r$ is chosen to reflect the conditions in Ref. \cite{BCWDyn}.   We see the increasing damping and natural frequency with rising $\xbar{W}$ and the dc response suppression appearing near $\Wpk$. For $\xbar{W} > W_{opt}$, the frequency of the relaxation oscillation moves back towards $\omega = 0$, as expected from the three-level model.  Near $W = W_{max}$, the transfer function no longer has a resonance as it monotonically decreases from its maximum at $\omega = 0$. 

We show the effect of the repumping ratio $r$ in Fig. \ref{fig:vsr}, where $\vec{\alpha}_+ = \vec{0}$, $\delta' = 0$ and $\xbar{W} = \Wpk$. The trends of lower damping and a lower natural frequency as $r \rightarrow 0$ are clearly visible, as expected from the three-level model in Sec. II.  

We also note that $T_A$ has a $1/\omega$ roll-off for $\omega \gg \omega_{res}$, even with higher order derivatives in the equations for $\jmath_\perp(\omega)$ and $\jmath_z(\omega)$ that function as a low-pass to the response (analogous to the $\dddot{\jmath}_\perp$ term in Eqn. \ref{eqn:DiffEqResponse1}). However, modulation of the repumping rate out of each hyperfine ground state, as was done in Ref. \cite{BCWDyn}, puts higher order derivatives in the drive terms as well.  The result is a drive that increases with a higher power of the modulation frequency $\omega$, partially balancing the higher order low-pass filtering.  Thus, by modulating the repumping rate out of each hyperfine state, the amplitude transfer function $T_A$ retains $1/\omega$ modulation frequency dependence of the three-level model in Sec. II.

The response as a function of $\delta'$ also qualitatively agrees with the simple picture put forward in Sec. II, as shown in Fig. \ref{fig:vsDelta}. For $\delta' > 0$ around $\delta' = 0$, we see a lower maximum $T_A$ consistent with heavier damping.  When $\delta' < 0$, the amplitude of the relaxation oscillations increase as the system becomes less damped.  But $\delta'$ continues to decrease, the full model shows a divergence in $T_A$ where the system becomes unstable with no steady-state solutions. In the unstable regime, we do not plot the transfer function and show a red shaded region in Fig. \ref{fig:vsDelta}b and \ref{fig:vsDelta}c. This instability is consistent with our inability to achieve steady-state superradiance experimentally at detuning $\delta' < -0.1$ \cite{BCWDyn}. The reduction in $\omega_{res}$ with increasing $\delta'$ is a result of maintaining the repumping $\xbar{W} = W_{opt}$, which reduces $\xbar{W}$ at large detunings and affects the natural frequency. 

\begin{figure}
\includegraphics[width=3.in]{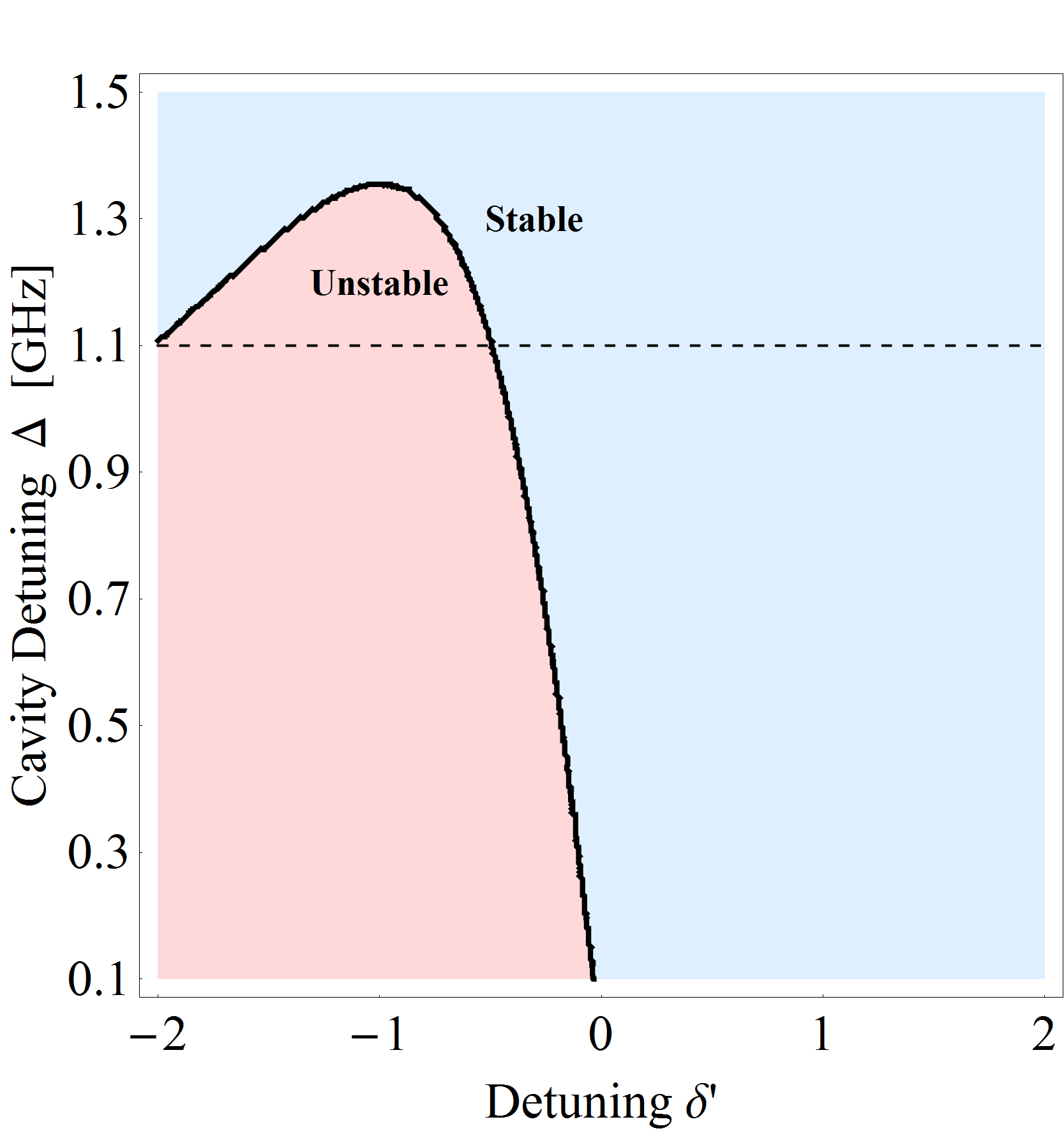}
\caption{Stability diagram for the full model of a superradiant Raman laser in $^{87}$ Rb, plotting the region any of the real parts of the poles of the $\jmath_\perp$ solution are positive. When any pole becomes positive, the system is unstable and has no steady-state solutions. The stability regions are shown versus the detuning of the cavity from the emission frequency $\delta'$ and the detuning of the bare cavity frequency from the atomic frequency $\Delta$. The critical contour (bold) marks where the pole changes sign. The dashed line indicates the detuning $\Delta$ of experimental work (Refs. \cite{BCW12, BCWDyn}). For the calculation we use $NC\gamma = 10^{-4}$ s$^{-1}$, $\xbar{W} = W_{opt}$, and $r = 0.71$. }
\label{fig:FullStabilityDiagram}
\end{figure}

We also use our linear response model to theoretically predict the stability diagram for the full $^{87}$Rb Raman laser system. We examine the poles of the solution for $\jmath_\perp$ as a function of $\delta'$, the detuning of the dressed cavity resonance frequency from the emission frequency and $\Delta$, the detuning of the bare cavity resonance frequency from the atomic lasing transition \ket{g} to \ket{i} = \ket{F'=2, m_f = \pm 1}.  We plot the regions of stability in Fig. \ref{fig:FullStabilityDiagram}, which is analogous to Fig. \ref{fig:TwoLevelStability} in Sec. II.  However, here the physical parameter $\Delta$ controls $\vec{\alpha}_+$, which roughly scales like $1/\Delta$ (we assume $\Delta$ remains large enough such that the system is well described by the dispersive tuning approximation).  Future experiments may benefit from working with larger detuning $\Delta$. However in the standing-wave geometry of Ref. \cite{BCWDyn},  the improved stability would come at the expense of increased inhomogeneous ac Stark shifts from the dressing laser.  At fixed scattering rate $\gamma$, the ac Stark shift increases linearly with $\Delta$.

\begin{figure}
\includegraphics[width=3.3in]{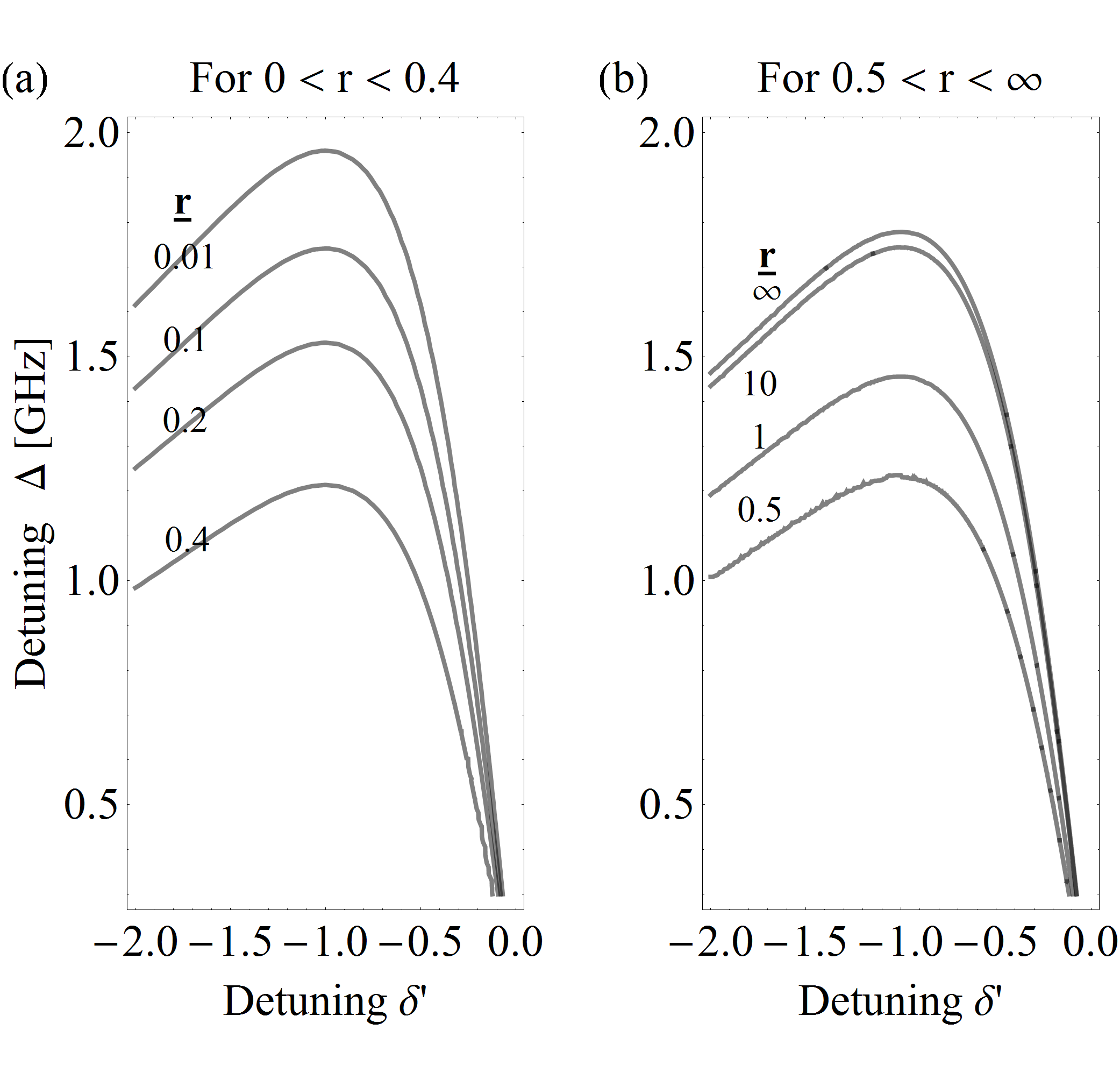}
\caption{Critical stability conditions for variable values of the repumping ratio $r$.  Each line shows the contour as a function of $\Delta$ and $\delta'$ separating stable lasing from unstable.  The unstable region is defined as any set of parameters that results in a positive value for the real part of any pole of the $J_\perp$ response solution. The stability conditions change as a function of the repumping ratio $r$. (a) As $r$ increases from 0, the unstable region gets smaller until it reaches some value between 0.4 and 0.5, after which (b) the unstable region grows to its asymptotic value.}
\label{fig:FullStabilityVs_r}
\end{figure}
\begin{figure}
\includegraphics[width=3.3in]{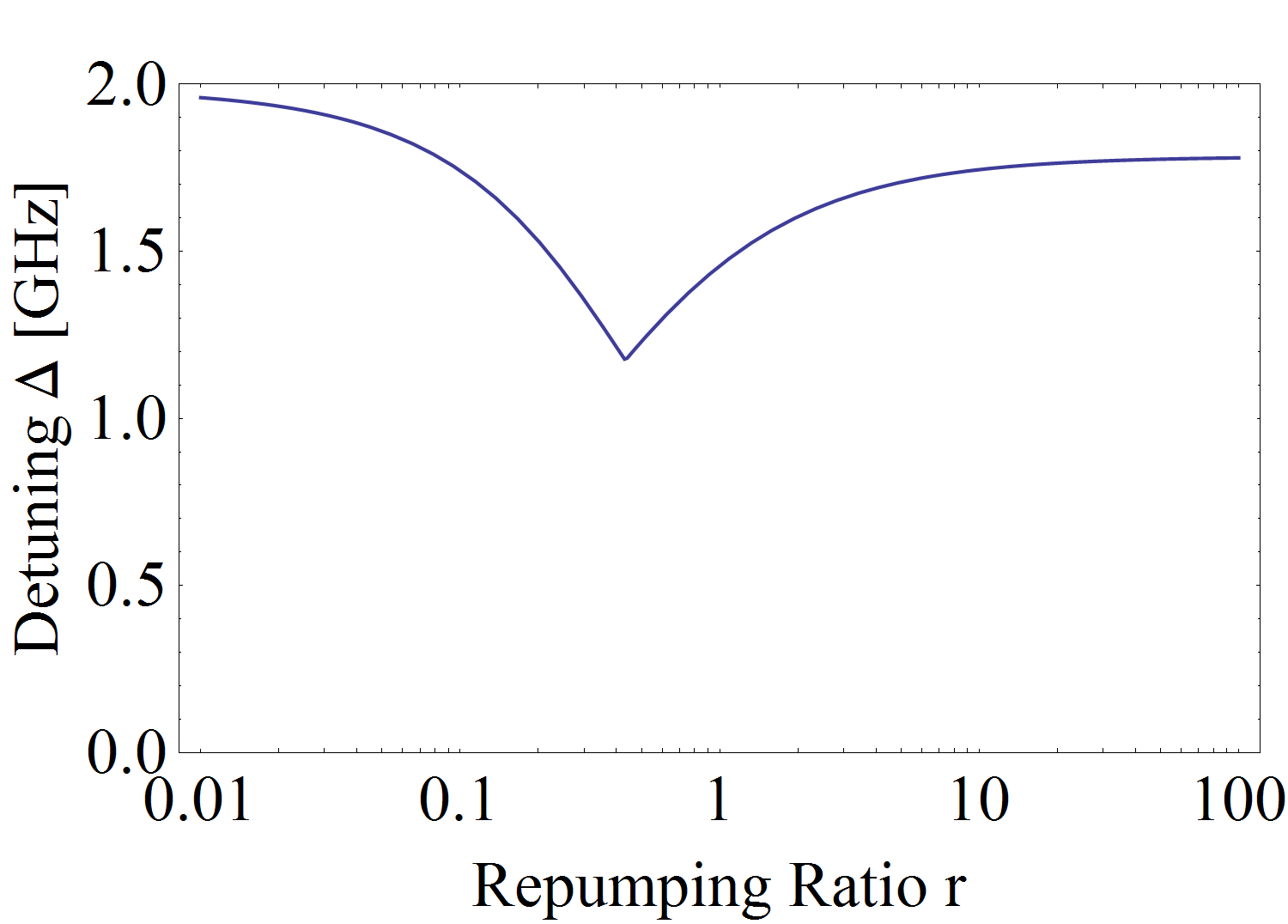}
\caption{The value of $\Delta$ for the critical contour versus $r$, assuming $\delta' = -1$. Lower $\Delta$ indicates that more of the parameter space has stable, steady-state solutions. Here we assume $NC\gamma = 10^{-4}$ s$^{-1}$ and $\xbar{W} = W_{opt}$.}
\label{fig:poleValueVsr}
\end{figure}

Repumping the atoms through multiple grounds states, quantified by the $r$ parameter, has a larger impact on the stability digram in this full model than on three level model in Sec. II. To study the effect repumping through the multiple ground states of $^{87}$Rb has on the stability of the laser amplitude, we follow the contour of the stability diagram for different values of $r$, shown in Fig. \ref{fig:FullStabilityVs_r}.  The figure is separated into two parts because the contour does not change monotonically.  In part (a), $r$ is low, indicating much of the population building up outside of the lasing levels, and the stable region grows in size as the repumping becomes more efficient.  However, as $r$ continues to grow, the contour asymptotes to an unstable region about the same size as if $r = 0.1$. In Fig. \ref{fig:poleValueVsr}, we plot the value of $\Delta$ for the critical contour, holding $\delta' = 1$, indicating that the largest stable region occurs when $r \approx 0.45$.  Here the cavity shift caused by atoms accumulating in the other hyperfine states acts to partially balance the shift from atoms in the \ket{g} and \ket{e} states, enhancing the amplitude stability.

\section{V. Conclusion}

We have developed a minimal model for a steady-state, superradiant laser that includes key features of observed in recent experiments using $^{87}$Rb \cite{BCW12, BCWDyn, BCWHybrid, WCB12}.  The model describes the reduction in the laser output power with the repumping ratio $r$, the cavity-atomic transition detuning $\delta$, and an additional source of decoherence, such as caused by Rayleigh scattering $\Gamma_R$. 
The model also describes the observed laser amplitude stability and provides a framework to understand the contributions of repumping and cavity tuning to the amplitude stability \cite{BCWDyn}. The explicit elimination of an intermediate excited state in our Raman laser theory shows that a Raman laser can serve as a good physics model for lasers operating deep into the bad-cavity regime.  The adiabatic elimination also reveals the source of the crucial atomic and cavity frequency tunings that can play a key role in the amplitude stability of Raman lasers, both in the bad-cavity \cite{BCWDyn} and good-cavity \cite{PhysRevLett.107.063904} regimes.

In addition to explaining experimental observations in previous work, this paper serves as a guide for the design of other cold-atom lasers and superradiant light sources that utilize nearly-forbidden optical transitions \cite{MYC09, CHE09}.  Our minimal model includes a multi-step repumping process and shows the path to adding more energy levels or repumping steps as required for realistic experimental systems.  Many of the results here do not assume a good-cavity or bad-cavity laser, making them general results that can be followed until simplified expressions based on a particular laser regime are required.

In general, superradiant laser designs should strive to eliminate sources of decoherence, such as Rayleigh scattering or differential ac Stark shifts from repumping light, while maintaining efficient repumping that avoids accumulation of population outside the atomic energy levels of the lasing transition. The steady-state and amplitude stability properties of cold-atom lasers can be significantly modified by their repumping scheme.  

Future designs may also apply optical dressing techniques to induce decay of the excited state \cite{WCB12, BCWHybrid}. In such Raman systems, the dressing of the cavity-mode can provide positive or negative feedback for stabilizing the output power of the laser.  The dressed cavity mode also can pull the laser emission frequency, serving as an amplitude noise to phase noise conversion mechanism. Future theoretical and experimental work, beyond the scope of this paper, can extend the linear response theory presented here to incorporate quantum noise in the repumping process.    Cavity frequency pulling and quantum noise in the dressing of the cavity mode are possible sources of the laser linewidth broadening observed in Ref. \cite{BCW12}, where the observed linewidth exceeded the simple Schawlow-Townes prediction \cite{MYC09}. 

\subsection{Acknowledgements}
The authors thank Murray Holland and Steven Cundiff for helpful conversations. All authors acknowledge financial support from DARPA QuASAR, ARO, NSF PFC, and NIST.  J.G.B. acknowledges support from NSF GRF, Z.C. acknowledges support from A*STAR Singapore, and K.C.C. acknowledges support from NDSEG. This material is based upon work supported by the National Science Foundation under Grant Number 1125844. 

\section{Appendix A: Details of the three level model}
\subsection{I. Liouvillian operators}

Here we give the individual Liouvillian terms present in the master equation of the three-level model, Eqn. \ref{eqn:masterEquation}, in Sec. II.  The Liouvillians give the dissipation associated with decay of the cavity mode,  spontaneous decay from \ket{e} to \ket{g},  spontaneous decay from \ket{3} to \ket{e}, Rayleigh scattering from state $\ket{g}$, and repumping from \ket{g} to \ket{3}, respectively.
 
\begin{equation}
\mathcal{L}_c[\hat{\rho}] = -\frac{\kappa}{2}(\hat{c}^\dag\hat{c}\hat{\rho} + \hat{\rho}\hat{c}^\dag\hat{c} -2\hat{c}\hat{\rho}\hat{c}^\dag) 
\end{equation}

\begin{equation} \mathcal{L}_{eg}[\hat{\rho}] = -\frac{\gamma}{2}\sum^N_{q=1}(\hat{\sigma}^{(q)}_{eg}\hat{\sigma}^{(q)}_{ge}\hat{\rho} + \hat{\rho}\hat{\sigma}^{(q)}_{eg}\hat{\sigma}^{(q)}_{ge} -2\hat{\sigma}^{(q)}_{ge}\hat{\rho}\hat{\sigma}^{(q)}_{eg})
\end{equation}

\begin{equation}
\mathcal{L}_{3e}[\hat{\rho}] = -\frac{\Gt}{2}\sum^N_{q=1}(\hat{\sigma}^{(q)}_{3e}\hat{\sigma}^{(q)}_{e3}\hat{\rho} + \hat{\rho}\hat{\sigma}^{(q)}_{3e}\hat{\sigma}^{(q)}_{e3} - 2\hat{\sigma}^{(q)}_{e3}\hat{\rho}\hat{\sigma}^{(q)}_{3e})
\end{equation}

\begin{equation}
\mathcal{L}_{R}[\hat{\rho}] = \frac{\Gamma_{R}}{4}\sum^N_{q=1}((\hat{\sigma}^{(q)}_{ee}-\hat{\sigma}^{(q)}_{gg}) \hat{\rho} (\hat{\sigma}^{(q)}_{ee}-\hat{\sigma}^{(q)}_{gg}) - (\hat{\sigma}^{(q)}_{ee}+\hat{\sigma}^{(q)}_{gg})\hat{\rho}).
\end{equation}

\begin{equation}
\mathcal{L}_{g3}[\hat{\rho}] = -\frac{W}{2}\sum^N_{q=1}(\hat{\sigma}^{(q)}_{g3}\hat{\sigma}^{(q)}_{3g}\hat{\rho} + \hat{\rho}\hat{\sigma}^{(q)}_{g3}\hat{\sigma}^{(q)}_{3g} -2\hat{\sigma}^{(q)}_{3g}\hat{\rho}\hat{\sigma}^{(q)}_{g3})
\end{equation}

\subsection{II. Full expressions for the three-level model linear response theory}

The full expressions for the coefficients in the three-level response equations Eqns. \ref{eqn:DiffEqResponse1} and \ref{eqn:DiffEqResponse2}:
\begin{align}\nonumber
\gamma_0 = \frac{r}{2\zeta} \biggl( (2N\Gamma_c-\Wp&) + 2\W r - \Gamma_R \\
&  +h(\delta)(4\gamma_\perp+\W(1+2r) )\biggr)
\end{align}

\begin{equation}
\omega_0^2 = \frac{-r(1+2r)\W}{\zeta} \biggl( (\Wp-N\Gamma_c)-\Wp h(\delta)  \biggr)
\end{equation}

\begin{equation}
\beta = \frac{(1+2r)}{W\zeta}
\end{equation}

\noindent where the denominator factor $\zeta= 2r h(\delta) + (1+r)(1+2r)$ and $h(\delta) = 2\alpha\delta\left(\frac{N}{1+\delta^2}-\frac{\Wp}{C\gamma} \right) $.

The drive terms are
\begin{equation}
D_{\perp,z}(\omega)  = D_{0,\perp,z} + i \omega D_{1,\perp,z} - \omega^2 D_{2,\perp,z}
\end{equation}

\noindent where the coefficients are 
\begin{equation}
D_{0,\perp} = \frac{-r(1+2r) \W}{2\zeta} \left( \W+\Wp-N\Gamma_c-\Gamma_R h(\delta) \right)
\end{equation}

\begin{equation}
D_{1,\perp} = - \frac{\W(1+r)(1+2r) + 2r\left(\Wp-N\Gamma_c-\Gamma_R h(\delta)\right)/2}{2\zeta} 
\end{equation}

\begin{equation}
D_{2,\perp} = - \frac{1+2r}{2\zeta} 
\end{equation}

\begin{equation}
D_{0,z} = \left(\frac{\W}{\Wp}\right) \frac{r(1+2r) \W (N\Gamma_c-\Wp)}{\zeta} 
\end{equation}

\begin{equation}
D_{1,z} = \left(\frac{\W}{\Wp}\right)\frac{r(3+2r) (N\Gamma_c-\Wp)}{\zeta} 
\end{equation}

\begin{equation}
D_{2,z} = \left(\frac{-2r}{\Wp}\right)\frac{N\Gamma_c-\Wp}{\zeta} 
\end{equation}

\subsection{III. Interesting limiting cases of three level solution}
Perfect repumping, on resonance:

\begin{equation}
\omega_0^2 = \W(NC\gamma-\Wp)
\end{equation}

\begin{equation}
\gamma_0 = \W/2
\end{equation}

\begin{equation}
\beta = 0
\end{equation}

\begin{equation}
D_\perp(\omega) = \frac{\W}{2}(NC\gamma-\W-\Wp-i\omega)
\end{equation}

\begin{equation}
D_z(\omega) =\left(\frac{\W}{\Wp}\right) (NC\gamma-\Wp)(\W+i\omega)
\end{equation}
\noindent Perfect repumping, with detuning:

\begin{equation}
\omega_0^2 = \W \left(N\Gamma_c-\Wp \right) \left(1+2\alpha \delta \frac{\Wp}{C\gamma}\right) 
\end{equation}

\begin{equation}
\gamma_0 = \frac{\W}{2} \left(1+h(\delta)\right)
\end{equation}

\begin{equation}
\beta = 0
\end{equation}

\begin{equation}
D_\perp(\omega) = \frac{\W}{2}\left(N\Gamma_c-\W-\Wp-\Gamma_R h(\delta) - i\omega \right)
\end{equation}

\begin{equation}
D_z(\omega) =\left(\frac{\W}{\Wp}\right) (N\Gamma_c-\Wp)(\W+i\omega)
\end{equation}
\noindent Imperfect repumping, on resonance:

\begin{equation}
\omega_0^2 = \left(\frac{r}{1+r}\right) \W \left(N\Gamma_c-\Wp \right)
\end{equation}

\begin{equation}
\gamma_0 = \left(\frac{r}{(1+r)(1+2r)}\right)(NC\gamma+(r+1/2)\W-\Wp)
\end{equation}

\begin{equation}
\beta = \frac{1}{\W(1+r)}
\end{equation}

\begin{align}
\begin{split}
D_\perp(\omega) &= \frac{\omega^2}{2(1+r)} \\
&+ \frac{i\omega}{2}\left(\frac{2r}{(1+r)(1+2r)}(NC\gamma-\Wp)-\W\right) \\ 
&+ \frac{r\W}{2(1+r)}(NC\gamma-\Wp-W) 
\end{split}
\end{align}

\begin{align}
\begin{split}
D_z(\omega) &= -\omega^2 \frac{2r(NC\gamma-\Wp)}{\Wp(1+r)(1+2r)} \\ 
&+ i \omega \left(\frac{\W}{\Wp}\right) \left(\frac{r(3+2r) (NC\gamma-\Wp) }{(1+r)(1+2r)}\right) \\
& +  \left(\frac{\W}{\Wp}\right)(NC\gamma-\Wp)
\end{split}
\end{align}

\section{Appendix B: Details of Raman laser model}

\subsection{I. Adiabatic elimination of the optically excited state} 

Here we explicitly derive the adiabatic elimination of an intermediate, optically excited state of a cold atom Raman laser described in Sec. III and Fig. \ref{fig:RamanEnLvl}.  The result is a system of equations describing the laser, Eqns. \ref{eqn:labframeCdot}-\ref{eqn:labframeJzdot}.

The Louivillian $\mathcal{L}[\hat{\rho}] = \mathcal{L}_c[\hat{\rho}]+\mathcal{L}_{ik}[\hat{\rho}] + \mathcal{L}_{ge}[\hat{\rho}] $ includes the cavity decay term $\mathcal{L}_c[\hat{\rho}]$, the spontaneous emission terms from state \ket{\mathrm{i}}

\begin{equation}
\mathcal{L}_{ik}[\hat{\rho}] = -\frac{\Gamma}{2}\sum^N_{q=1}\sum_{k=e,g}(\hat{\sigma}^{(q)}_{ik}\hat{\sigma}^{(q)}_{ki}\hat{\rho} + \hat{\rho}\hat{\sigma}^{(q)}_{ik}\hat{\sigma}^{(q)}_{ki} -2\hat{\sigma}^{(q)}_{ki}\hat{\rho}\hat{\sigma}^{(q)}_{ik}),
\end{equation}

\noindent and an incoherent repumping term that looks like spontaneous decay from \ket{g} to \ket{e} 

\begin{equation}
\mathcal{L}_{ge}[\hat{\rho}] = -\frac{W}{2}\sum^N_{q=1}(\hat{\sigma}^{(q)}_{ge}\hat{\sigma}^{(q)}_{eg}\hat{\rho} + \hat{\rho}\hat{\sigma}^{(q)}_{ge}\hat{\sigma}^{(q)}_{eg} -2\hat{\sigma}^{(q)}_{eg}\hat{\rho}\hat{\sigma}^{(q)}_{ge}).
\end{equation}

\noindent  As in Sec. II, we assume we are able to factorize the expectation values $\expec{\hat{c}\hat{\sigma}_{ie}} = \mathcal{C}\sigma_{ie}$, $\expec{\hat{c}\hat{\sigma}_{ii}} = \mathcal{C}\sigma_{ii}$, $\expec{\hat{c}\hat{\sigma}_{gg}} = \mathcal{C}\sigma_{gg}$, $\expec{\hat{c}\hat{\sigma}_{eg}} = \mathcal{C}\sigma_{eg}$, $\expec{\hat{c}\hat{\sigma}_{ig}} = \mathcal{C}\sigma_{ig}$, and  $\expec{\hat{c}\hat{\sigma}_{ie}} = \mathcal{C}\sigma_{ie}$.

Applying these assumptions to the master equation results in the equations of motion

\begin{align}
\label{eqn:RamanCpretransform}
\dot{\mathcal{C}} =&  -(\kappa/2 + i \omega_c)\mathcal{C} - i g J_{gi} \\
\dot{J}_{ge} =& -(W/2+i\omega_{eg}) J_{ge} - i \frac{\Omega_d}{2}e^{i\omega_d t} J_{gi} + i g \mathcal{C} J_{ie} \\
\begin{split}
\dot{J}_{gi} =& -\left(\frac{\Gamma+W}{2} + i\omega_{ig}\right)J_{gi} \\&-  i \frac{\Omega_d}{2}e^{-i\omega_d t} J_{ge} + ig \mathcal{C} (N_i - N_g)
\end{split}\\
\begin{split}
\dot{J}_{ei} =& -\left(\frac{\Gamma+W}{2} + i\omega_{ie}\right) J_{ei} \\&+ i \frac{\Omega_d}{2}e^{-i\omega_d t} (N_i - N_e) - i g \mathcal{C} J_{eg}
\end{split}\\
\begin{split}
\dot{J}_z =& \frac{1}{2}(\dot{N}_e-\dot{N}_g) = W (N/2 - J_z) \\& -i \frac{\Omega_d}{2} (\sigma_{ei}e^{i\omega_d t} -J_{ie}e^{-i\omega_d t}) + ig(\mathcal{C}^*J_{gi}-\mathcal{C}J_{ig}).
\label{egn:RamanSigZprerransform}
\end{split}
\end{align}

\noindent Here we identity the relevant transverse atomic decay rate $\gamma_\perp = W/2$.

The equation for $\dot{J}_z$ assumes only a negligible fraction of the atomic ensemble resides in \ket{\mathrm{i}}, an assumption we justify shortly.  For convenience, we go into a rotating frame (often called the natural frame \cite{BPM07}) defined by the transformation of variables:

\begin{gather}
\tilde{\mathcal{C}} = \mathcal{C} e^{i \omega_c t} \\ 
\tilde{J}_{ge} =  J_{ge}e^{i\omega_{eg} t}\\
\tilde{J}_{gi} = J_{gi}e^{i(\omega_c + \delta_0/2) t}\\
\tilde{J}_{ei} = J_{ei}e^{i(\omega_d - \delta_0/2) t}\\
\end{gather}

\noindent where $\delta_0$ is the two-photon detuning $\delta_0 = \omega_d-\omega_c+\omega_{eg}$.  With these substitutions, Eqns. \ref{eqn:RamanCpretransform} - \ref{egn:RamanSigZprerransform} become

\begin{gather}
\dot{\tilde{\mathcal{C}}} =  -(\kappa/2) \tilde{\mathcal{C}} - i g \tilde{J}_{gi} e^{-i\delta_0 t / 2} \\
\dot{\tilde{J}}_{ge} = -\gamma_\perp \tilde{J}_{ge} - i \frac{\Omega_d}{2}e^{i\delta_0 t/2} \tilde{J}_{gi} + i g \tilde{\mathcal{C}} \tilde{J}_{ie}e^{i\delta_0 t/2}
\end{gather}
\begin{align}
\begin{split}
\dot{\tilde{J}}_{gi} =& (i \Delta - \frac{\Gamma+W}{2}) \tilde{J}_{gi} -  i \frac{\Omega_d}{2}e^{-i\delta_0 t/2} \tilde{J}_{ge} \\&+ ig\tilde{\mathcal{C}}e^{i\delta_0t/2}(N_i - N_g) 
\end{split}
\end{align}
\begin{align}
\begin{split}
\dot{\tilde{J}}_{ei} =& (i \Delta - \frac{\Gamma+W}{2}) \tilde{J}_{ei} + i \frac{\Omega_d}{2}e^{-i\delta_0 t/2} (N_i - N_e) \\
&- i g \tilde{\mathcal{C}} e^{i\delta_0 t/2} \tilde{J}_{eg}
\end{split}
\end{align}
\begin{align}
\begin{split}
\dot{J}_z =& W (N/2 - J_z) -i \frac{\Omega_d}{2} (\tilde{J}_{ei}e^{i \delta_0 t/2} -\tilde{J}_{ie}e^{-i \delta_0 t/2})  \\&+ ig(\tilde{\mathcal{C}}^*\tilde{J}_{gi}e^{-i \delta_0 t/2}-\tilde{\mathcal{C}}\tilde{J}_{ig}e^{i \delta_0 t/2})
\end{split}
\end{align}
\vspace{3mm}

\noindent Here $\Delta= \Delta_d + (\delta_0/2)$ is also equivalent to the average detuning of the Raman dressing laser $\Delta_d$ and the cavity mode $\Delta_c$ from their respective optical atomic transitions.

To reduce these equations to those of an effective two-level system coupled to a cavity field, we assume that we can adiabatically eliminate the collective coherences $\tilde{J}_{gi}, \tilde{J}_{ei}$ and that the population of the intermediate state is small $N_i \ll N_g$, $N_e$.  These assumptions are justified due to large detuning $\Delta \gg \Gamma, \gamma_\perp, \delta_0$. The adiabatic elimination of the coherence proceeds as follows \cite{BPM07}: by examining the form of the equations for $\dot{\tilde{J}}_{gi}$ and $\dot{\tilde{J}}_{ei}$, we expect that each one can be written as the sum of a term rapidly oscillating at frequency $\Delta$, and a term varying on the timescale of the population dynamics, much more slowly than $1/\Delta$. By averaging over a timescale long compared to the rapid oscillation, but short compared to the population dynamics, we essentially perform a coarse graining approximation and are left with only slowly varying terms.  The derivatives of these coarse-grained collective amplitudes are negligible. Here we consider small fluctuations about steady-state values at frequency $\omega$, so the approximation will be valid when $\Delta \gg \omega$.  Thus, to a good approximation for the cases considered here, the derivatives $\dot{\tilde{J}}_{gi}, \dot{\tilde{J}}_{ei}$  can be set to zero.  We then solve for $\tilde{J}_{gi}$ and $\tilde{J}_{ei}$ as 


\begin{gather}
\tilde{J}_{gi} \approx \frac{\Omega_d}{2 \Delta} e^{-i \delta_0 t/2} \tilde{J}_{ge} + \frac{g}{\Delta}e^{i \delta_0 t/2} \tilde{\mathcal{C}} N_g \\
\tilde{J}_{ei} \approx \frac{\Omega_d}{2 \Delta} e^{i \delta_0 t/2} N_e + \frac{g}{\Delta}e^{-i \delta_0 t/2} \tilde{\mathcal{C}} \tilde{J}_{eg} 
\end{gather}

\noindent where we have approximated $i\Delta + \frac{\Gamma +W}{2} \approx i\Delta$.  After including these simplifications and transforming back to the original frame, we arrive at Eqns. \ref{eqn:labframeCdot}-\ref{eqn:labframeJzdot} in the main text.

\subsection{II. Steady-state emission frequency of the Raman laser}

We find the steady-state cold atom Raman laser frequency from Sec. III by assuming the laser is oscillating at frequency $\omega_\gamma$, so $\mathcal{C} = \breve{\mathcal{C}} e^{-i \omega_\gamma t}$ and $J_{eg} = \breve{J_{eg}}e^{-i (\omega_\gamma-\omega_d) t}$.  Substituting in Eqns. \ref{eqn:labframeCdot} - \ref{eqn:labframeJzdot} gives 

\begin{gather}
\dot{\breve{\mathcal{C}}} = (-\kappa/2-i\delta)\breve{\mathcal{C}} - i g_2 \breve{J}_{ge} \label{eqn:RamanCAmplitude}\\
\dot{\breve{J}}_{ge} =\left(-\gamma_\perp - i(\omega_{eg} -\omega_{ac} +\omega_d -\omega_\gamma)\right)\breve{J}_{ge} + i 2 g_2 J_z \breve{\mathcal{C}}  \label{eqn:RamanJegAmplitude} \\
\dot{J}_z =W(N/2-J_z) + i g_2 (\breve{\mathcal{C}}^* \breve{J}_{ge} - \breve{\mathcal{C}} \breve{J}_{eg})
\label{eqn:RamanJzAmplitude}
\end{gather}

\noindent where $\delta$ is the detuning of the emission frequency from the dressed cavity mode 

\begin{equation}
\delta = \omega_D -\omega_\gamma \,\,.
\label{eqn:dressedcavitydetuning}
\end{equation}

\noindent The steady-state emission frequency $\omega_\gamma$ is constrained by the condition that $J_z$ must be real, and following the procedure in Sec. II, we arrive at the  the laser oscillation frequency

\begin{equation}
\omega_\gamma = \frac{2\gamma_\perp}{2\gamma_\perp+\kappa}\omega_D + \frac{\kappa}{2\gamma_\perp+\kappa}(\omega_{eg} + \omega_d -\omega_{ac})\,\, .
\end{equation}

\noindent Note the insensitivity of the oscillation frequency to changes in the cavity frequency in the bad-cavity limit where $\kappa \gg W = 2\gamma_\perp$. 

\section{Appendix C: Details for the $^{87}$Rb Full Model} 

\subsection{I. Repumping scheme}
We begin the description of our model for a cold atom laser in $^{87}$Rb with the details of the repumping process. The equations to describe the repumping are arrived at after adiabatic elimination of the optically excited states $\ket{5^2P_{3/2}, F'=2, m_F}$ through which the Raman transitions for repumping are driven.  However, unlike in the Sec. III, the scattered photons lack a resonant cavity mode, so the scattering is presumed to be primarily into free space (i.e. non-cavity) modes.

The relevant set of Rabi frequencies describing the resonant repumper laser coupling the ground state \ket{5^2S_{1/2}, F, m_F} state to an optically excited state \ket{5^2P_{3/2},F'=2, m_F' } is given by the dipole matrix element between the states as

\begin{equation}
\Omega_{F, m_F, 2 , m_F'} = \left | \bra{5^2P_{3/2},2, m_F } \vec{d}\cdot \vec{E}_F\ket{5^2S_{1/2}, F, m_F} \right |/\hbar
\end{equation}


\noindent   where $\vec{d}$ is the atomic dipole moment operator and the electric field of the two repumping lasers are $\vec{E}_1$ and $\vec{E}_2$.  If an atom is in the excited state \ket{5^2P_{3/2},F'=2, m_F' } then it spontaneously decays to the ground state \ket{5^2S_{1/2}, F, m_F} with fractional probability given by the branching ratio 

\begin{equation}
B_{F, m_F, F', m_F'} =|\langle F \,m_f | F'\,1\, m'_f\, p \rangle|^2  \,\, ,
\end{equation}

\noindent where $p$ labels the polarization of the emitted light ($\sigma^+=-1, \pi = 0, \sigma^- = +1$ and $\sum_{F, m_F}B_{F, m_F, F', m_F'} =1$.

The repumping rate $W'$ in our model is calculated as the resonant, unsaturated scattering rate

\begin{equation}
W' = \frac{\Omega^2_{1, 0, 2, 0 }}{2 \Gamma_{D2}}(1-B_{1,0,2,0})
\end{equation}

\noindent where $\Gamma_{D2}$ is the D2 excited state decay rate $\Gamma_{D2}/2\pi = 6.07$ MHz. Note that the rate $W'$ is the scattering rate out of the ground state, and does not include Rayleigh scattering into free space $\Gamma_R = \frac{\Omega^2_{1, 0, 2, 0 }}{2 \Gamma}B_{1,0,2,0}$ which causes the scattering atom to collapse into $\ket{g}$ due the optically thin nature of the atomic ensemble along nearly all directions other than the cavity mode.  Since  $\Gamma_R$ scales with $\frac{\Omega^2_{1, 0, 2, 0 }}{2 \Gamma}$ as does $W'$, we will group both together into a single rate $W = \frac{\Omega^2_{1, 0, 2, 0 }}{2 \Gamma_{D2}}$, and distinguish the two rates with branching ratios in our equations for the population equations.

\subsection{II. Reduced optical Bloch equations}


Including the coherent dynamics of the effective two-level system adds the coherence $\Jperp$ driving the population from \ket{e} to \ket{g} as was seen in Sec. III, Eqn. \ref{eqn:JzRamanOnly}

\begin{equation}
\dot{N}_{e,g} = \mp\frac{C\gamma}{1+\delta'^2}J_\perp^2 \,\,.
\end{equation}

\noindent Here $\delta'$ is the detuning of the dressed cavity mode from the emitted light frequency, normalized by $\kappa/2$, as in Eqn. \ref{eqn:dressedcavitydetuning}. In the subsequent equations, we neglect the single particle scattering from \ket{e} to \ket{g} at rate $\gamma$ as it is much less than the collective emission rate.

With these terms, we can write the reduced optical Bloch equations for the ground state populations as

\begin{widetext}
\begin{equation}
\frac{\mathrm{d}N_{F, m_F}}{\mathrm{d} t} = \frac{W}{\Omega^2_{1, 0, 2, 0 }}  \sum_{F' = 1}^2  \sum_{m_F' = -F'}^{F'}  \left (B_{F, m_F, 2, m_F'} -\delta_{F, F'}, \delta_{m_F, m_F'}\right)\Omega^2_{F',m_F', 2, m_F' } N_{F', m_F' } + \frac{C \gamma}{1+\delta'^2} \Jperp \left( \delta_{F, 1}, \delta_{m_F, 0}  - \delta_{F, 2}, \delta_{m_F, 0}\right)
\end{equation}
\end{widetext}

\noindent where $\delta_{F,F'}$ is the Kronecker delta function. The sums have been reduced using the assumption that the repumping light is $\pi$-polarized.


We also must include the equation for the coherence, driven by the population inversion

\begin{equation}
\dot{\Jperp} = -\Goz \Jperp + \frac{C \gamma}{1+\delta'^2}(N_{2, 0}-N_{1, 0}) \Jperp
\label{eqn:coherence}
\end{equation}

\noindent which is analogous to Eqn. \ref{eqn:JperpRamanOnly} in Sec. III, except that here $W$ is the sum of the ground state repumping rate and the Rayleigh scattering rate,  where $W$ in Eqn. \ref{eqn:JperpRamanOnly} contains only the ground state repumping rate.

The repumping rates induced by the F2 repumper are parameterized by repumping ratio $r$ defined as

\begin{equation}
r = \frac{\Omega^2_{2, 1, 2, 1 }}{\Omega^2_{1, 0, 2, 0 }} \, .
\end{equation}

The normalized detuning $\delta'$ of the dressed cavity resonant frequency with the emitted light frequency $\omega_\gamma$ carries implicit dependence on the populations $N_{F, m_F}$ as derived in Sec. III

\begin{equation}
\delta' = 2 \left(\omega_c + \vec{\alpha} \cdot \vec{V} - \omega_\gamma \right)/\kappa \, .
\end{equation} 

\noindent Here $\vec{V}$ is a column vector of the populations in ground hyperfine states  

\begin{equation}
\vec{V} = \left(\begin{array}{l}
N_{2,2}\\
N_{2,1}\\
N_e\\
N_{2,-1}\\
N_{2,-2}\\
N_{1,1}\\
N_g\\
N_{1,-1}\\
\Jperp
\end{array}
\right)
\end{equation}

\noindent where  $N_e = N_{2, 0}$ and $N_g = N_{1, 0}$ are specially labeled to indicate their importance as the lasing levels.

The elements of the single-atom cavity tuning vector $\vec{\alpha}$ comes from the cavity dressing as derived from Eqn. \ref{eqn:dressedCavityFreqRamanOnly}.  There we see $\vec{\alpha}$ is set by the detuning of the cavity frequency and the detuning from the atomic transition frequencies $\omega_{F, F'}$  between the ground  \ket{5^2S_{1/2}, F} states and optically excited states \ket{5^2P_{1/2},F' } as

\begin{equation}
 \alpha_{F, m_F} =  \sum_{F'= 1}^{2} \frac{(2 g_{F, m_F, F', m_F+p})^2}{4 (\omega_{bcav} - \omega_{F, F'})}
\end{equation}

\noindent where $p=\pm1$ for the $\sigma^\pm$ polarized cavity mode, and the single-particle vacuum Rabi frequencies are evaluated for each transition.  For the quantization axis along the cavity axis, the $\sigma^+$ and $\sigma^-$ polarizations modes will shift in frequency by different amounts specified by two vectors $\vec{\alpha}_\pm$. However, the symmetry of the atomic population equations ensure that the populations are symmetric such that $N_{F, m_F}  =N_{F, -m_F}$.  Thus, only the shift of the $\sigma^+$ cavity mode needs to be considered. We use the dressing laser detuning $\Delta = +1.1$ GHz as was present in Refs. \cite{BCW12, BCWDyn}. The resulting cavity tuning vector is 

\begin{equation}
\vec{\alpha}_+ = 2 \pi (33.8~\mathrm{Hz} )\left(\begin{array}{c}
0 \\ 
0.0776 \\
0.1515 \\
0.222 \\
0.289\\
1.679\\
1\\
0.440\\
0\\
\end{array}  \right) \,\, .
\end{equation}

We find the steady-state solutions to the system of equations by setting $\mathrm{d}\vec{V}/\mathrm{d}t =0$ and solving for $\vec{V}$.  From the results, we can for the expressions for $\bar{J}_\perp^2$, $\bar{J}_z$, and $\bar{N}_{other}$ in the main text.

\end{document}